\documentclass[twocolumn]{aastex62}

\definecolor{purple}{rgb}{0.5,0,0.5}
\definecolor{darkgreen}{rgb}{0.0,0.5,0}
\definecolor{darkyellow}{rgb}{0.5,0.5,0}
\definecolor{violet}{rgb}{0.5,0,0.8}

\graphicspath{{./}{figures/}}

\date{\today}

\shorttitle{The BUFFALO HST Survey}
\shortauthors{Steinhardt et al.}

\begin{document}

\title{The BUFFALO HST Survey}

\correspondingauthor{Charles L. Steinhardt}
\email{steinhardt@nbi.ku.dk}

\author[0000-0003-3780-6801]{Charles L. Steinhardt}
\affiliation{Cosmic Dawn Center (DAWN)}
\affiliation{Niels Bohr Institute, University of Copenhagen, Lyngbyvej 2, Copenhagen \O~2100}

\author[0000-0003-1974-8732]{Mathilde Jauzac}
\affiliation{Centre for Extragalactic Astronomy, Durham University, South Road, Durham DH1 3LE, U.K.}
\affiliation{Institute for Computational Cosmology, Durham University, South Road, Durham DH1 3LE, U.K.}
\affiliation{Astrophysics and Cosmology Research Unit, School of Mathematical Sciences, University of KwaZulu-Natal, Durban 4041, South Africa}



\author[0000-0003-3108-9039]{Ana Acebron}
\affiliation{Physics Department, Ben-Gurion University of the Negev, P.O. Box 653, Be'er-Sheva 8410501, Israel}

\author{Hakim Atek}
\affiliation{Institut d'astrophysique de Paris, CNRS UMR7095, Sorbonne Universit\'e, 98bis Boulevard Arago, F-75014 Paris, France}

\author{Peter Capak}
\affiliation{IPAC, California Institute of Technology, 1200 E. California Blvd., Pasadena, CA, 91125}
\affiliation{Cosmic Dawn Center (DAWN)}

\author[0000-0002-2951-7519]{Iary Davidzon}
\affiliation{Cosmic Dawn Center (DAWN)}
\affiliation{Niels Bohr Institute, University of Copenhagen, Lyngbyvej 2, Copenhagen \O~2100}

\author{Dominique Eckert}
\affiliation{University of Geneva, Ch. d'Ecogia 16, CH-1290 Versoix Switzerland}

\author[0000-0002-6066-6707]{David Harvey}
\affiliation{Lorentz Institute, Leiden University, Niels Bohrweg 2, Leiden, NL-2333 CA, The Netherlands}

\author[0000-0002-6610-2048]{Anton M. Koekemoer}
\affiliation{Space Telescope Science Institute, 3700 San Martin Dr., Baltimore, MD 21218, USA}

\author[0000-0003-3021-8564]{Claudia D.P. Lagos} 
\affiliation{International Centre for Radio Astronomy Research (ICRAR), University of Western Australia, 35 Stirling Hwy, Crawley, WA 6009, Australia.}
\affiliation{ARC Centre of Excellence for All Sky Astrophysics in 3 Dimensions (ASTRO 3D)}
\affiliation{Cosmic Dawn Center (DAWN)}

\author[0000-0003-3266-2001]{Guillaume Mahler}
\affiliation{Department of Astronomy, University of Michigan, 1085 South University Ave, Ann Arbor, MI 48109, USA}

\author[0000-0001-7847-0393]{Mireia Montes}
\affiliation{School of Physics, University of New South Wales, Sydney, NSW 2052, Australia}

\author{Anna Niemiec}
\affiliation{Department of Astronomy, University of Michigan, 1085 South University Ave, Ann Arbor, MI 48109, USA}

\author[0000-0001-6342-9662]{Mario Nonino}
\affiliation{INAF-Trieste Astronomical Observatory}

\author[0000-0001-5851-6649]{P. A. Oesch}
\affiliation{Department of Astronomy, University of Geneva, 51 Ch. des Maillettes, 1290 Versoix, Switzerland}
\affiliation{International Associate, Cosmic Dawn Center (DAWN)}

\author{Johan Richard}
\affiliation{Univ Lyon, Univ Lyon1, Ens de Lyon, CNRS, Centre de Recherche Astrophysique de Lyon UMR5574, F-69230, Saint-Genis-Laval, France}

\author[0000-0003-1947-687X]{Steven A. Rodney}
\affiliation{University of South Carolina, Department of Physics \& Astronomy, Columbia, SC, USA}


\author[0000-0002-2395-4902]{Matthieu Schaller}
\affiliation{Leiden Observatory, Leiden University, P.O. Box 9513, 2300 RA Leiden, the Netherlands}

\author[0000-0002-7559-0864]{{Keren Sharon}}
\affiliation{Department of Astronomy, University of Michigan, 1085 South University Ave, Ann Arbor, MI 48109, USA}

\author[0000-0002-7756-4440]{Louis-Gregory Strolger}
\affiliation{Space Telescope Science Institute}
\affiliation{Johns Hopkins University}


\author[0000-0003-2718-8640]{Joseph Allingham}
\affiliation{Sydney Institute for Astronomy, School of Physics, A28, The University of Sydney, NSW 2006, Australia}

\author[0000-0003-3481-3491]{Adam Amara}
\affiliation{Institute of Cosmology \& Gravitation, University of Portsmouth, Dennis Sciama Building, Portsmouth, PO1 3FX}

\author[0000-0002-3196-5126]{Yannick Bah'{e}}
\affiliation{Leiden Observatory, Leiden University, P.O. Box 9513, 2300 RA Leiden, The Netherlands}

\author[0000-0002-5074-9998]{C\'eline B\oe hm}
\affiliation{University of Sydney, School of Physics, Sydney  NSW 2006, Australia}

\author[0000-0002-0974-5266]{Sownak Bose}
\affiliation{Center for Astrophysics \textbar Harvard \& Smithsonian, 60 Garden Street, Cambridge, MA 02138, USA}

\author[0000-0002-4989-2471]{Rychard J. Bouwens}
\affiliation{Leiden Observatory,   Leiden University, NL-2300 RA Leiden, Netherlands}

\author[0000-0002-7908-9284]{Larry D. Bradley}
\affiliation{Space Telescope Science Institute, 3700 San Martin Dr., Baltimore, MD 21218, USA}

\author[0000-0003-2680-005X]{Gabriel Brammer}
\affiliation{Cosmic Dawn Center (DAWN)}
\affiliation{Niels Bohr Institute, University of Copenhagen, Lyngbyvej 2, Copenhagen \O~2100}

\author{Tom Broadhurst}
\affiliation{Dept. of Physics, University. of the Basque Country}
\affiliation{Donostia Internaltional Physics Center, San Sebastian, Spain}
\affiliation{Ikerbasque Foundation }

\author[0000-0003-0776-4102]{Rodrigo Ca\~nas}
\affiliation{International Centre for Radio Astronomy Research, University of Western Australia, 35 Stirling Highway, Crawley, WA 6009, Australia}
\affiliation{ARC Centre of Excellence for All Sky Astrophysics in 3 Dimensions (ASTRO 3D)}

\author{Renyue Cen}
\affiliation{Department of Astrophysical Sciences, 4 Ivy Lane, Princeton, NJ 08544, USA}

\author{Benjamin Cl\'ement}
\affiliation{Institute of Physics, Laboratory of Astrophysics, Ecole Polytechnique Fédérale de Lausanne (EPFL), Observatoire de Sauverny, 1290 Versoix, Switzerland}

\author{Douglas Clowe}
\affiliation{Department of Physics \& Astronomy, Ohio University, Clippinger Labs 251B, Athens, OH 45701, USA}

\author[0000-0001-7410-7669]{Dan Coe}
\affiliation{Space Telescope Science Institute, 3700 San Martin Dr., Baltimore, MD 21218, USA}

\author[0000-0002-7898-7664]{Thomas Connor}
\affiliation{Jet Propulsion Laboratory, California Institute of Technology, Pasadena, CA 91109, USA}
\affiliation{The Observatories of the Carnegie Institution for Science, 813 Santa Barbara St., Pasadena, CA 91101, USA}

\author[0000-0003-4919-9017]{Behnam Darvish}
\affiliation{Cahill Center for Astrophysics, California Institute of Technology, 1216 East California Boulevard, Pasadena, CA 91125, USA}

\author[0000-0001-9065-3926]{Jose M. Diego}
\affiliation{Instituto de F\'isica de Cantabria (CSIC-UC).  Edificio Juan Jord\'a. Avda Los Castros s/n. 39005 Santander, Spain }

\author{Harald Ebeling}
\affiliation{Institute for Astronomy, University of Hawaii, 2680 Woodlawn Dr, Honolulu, HI 96822, USA}

\author[0000-0002-3398-6916]{A. C. Edge}
\affiliation{Centre for Extragalactic Astronomy, Durham University, South Road, Durham DH1 3LE, U.K.}

\author[0000-0003-1344-9475]{Eiichi Egami}
\affiliation{Steward Observatory, University of Arizona, 933 N. Cherry Avenue, Tucson, AZ 85721, USA}

\author[0000-0003-4117-8617]{Stefano Ettori}
\affiliation{INAF, Osservatorio di Astrofisica e Scienza dello Spazio, via Pietro Gobetti 93/3, 40129 Bologna, Italy}
\affiliation{INFN, Sezione di Bologna, viale Berti Pichat 6/2, I-40127 Bologna, Italy}

\author[0000-0002-9382-9832]{Andreas L. Faisst}
\affiliation{IPAC, California Institute of Technology, 1200 E. California Blvd., Pasadena, CA, 91125}

\author[0000-0003-1625-8009]{Brenda Frye}
\affiliation{Department of Astronomy/Steward Observatory, 933 North Cherry Ave., University of Arizona, Tucson AZ  85721}

\author[0000-0001-6278-032X]{Lukas J. Furtak}
\affiliation{Institut d'astrophysique de Paris, Sorbonne Universit\'{e}, CNRS UMR 7095, 98 bis bd Arago, 75014 Paris, France}

\author[0000-0002-4085-9165]{C. G\'omez-Guijarro}
\affiliation{Laboratoire AIM-Paris-Saclay, CEA/DSM-CNRS-Universit\'e Paris Diderot, Irfu/Service d’Astrophysique, CEA Saclay, Orme des Merisiers, F-91191 Gif-sur-Yvette, France}

\author[0000-0002-7868-9827]{J. D. Remolina Gonz\'{a}lez}
\affiliation{Department of Astronomy, University of Michigan, 1085 South University Ave, Ann Arbor, MI 48109, USA}

\author[0000-0002-0933-8601]{Anthony Gonzalez}
\affiliation{Department of Astronomy, University of Florida, 211 Bryant Space Center, Gainesville, FL 32611, USA}

\author[0000-0002-4391-6137]{Or Graur}
\affiliation{Center for Astrophysics \textbar Harvard \& Smithsonian, 60 Garden Street, Cambridge, MA 02138, USA}
\affiliation{Department of Astrophysics, American Museum of Natural History, New York, NY 10024, USA}
\affiliation{NSF Astronomy and Astrophysics Postdoctoral Fellow}

\author[0000-0003-3270-7644]{Daniel Gruen}
\affiliation{Kavli Institute for Particle Astrophysics and Cosmology, P. O. Box 2450, Stanford University, Stanford, CA 94305, USA}
\affiliation{Department of Physics, Stanford University, 382 Via Pueblo Mall, Stanford, CA 94305, USA}

\author[0000-0002-6066-6707]{David Harvey}
\affiliation{Lorentz Institute, Leiden University, Niels Bohrweg 2, Leiden, NL-2333 CA, The Netherlands}

\author{Hagan Hensley}
\affiliation{California Institute of Technology, 1200 East California Boulevard, Pasadena, CA 91125, USA}
\affiliation{Cosmic Dawn Center (DAWN)}

\author{Beryl Hovis-Afflerbach}
\affiliation{Division of Physics, Mathematics, and Astronomy, California Institute of Technology, Pasadena, CA 91125, USA}
\affiliation{Cosmic Dawn Center (DAWN)}

\author{Pascale Jablonka}
\affiliation{Institute of Physics, Laboratoire d’astrophysique, Ecole Polytechnique F\'ed\'erale de Lausanne (EPFL), Observatoire, 1290 Versoix, Switzerland}
\affiliation{GEPI, Observatoire de Paris, Universit\'e PSL, CNRS, Place Jules Janssen, F-92190, Meudon, France}

\author[0000-0001-8738-6011]{Saurabh~W. Jha}
\affiliation{Rutgers, the State University of New Jersey, 136 Frelinghuysen Road, Piscataway, NJ 08854, USA}

\author[0000-0002-9253-053X]{Eric Jullo}
\affiliation{Aix Marseille Université, CNRS, LAM (Laboratoire d'Astrophysique de Marseille) UMR 7326, 13388, Marseille, France}

\author[0000-0002-4616-4989]{Jean-Paul Kneib}
\affiliation{Institute of Physics, Laboratory of Astrophysics, Ecole Polytechnique F\'ed\'erale de Lausanne (EPFL), Observatoire de Sauverny, 1290 Versoix, Switzerland.}
\affiliation{Aix Marseille Université, CNRS, LAM (Laboratoire d'Astrophysique de Marseille) UMR 7326, 13388, Marseille, France}

\author[0000-0002-5588-9156]{Vasily Kokorev}
\affiliation{Cosmic Dawn Center (DAWN)}
\affiliation{Niels Bohr Institute, University of Copenhagen, Lyngbyvej 2, Copenhagen \O~2100}

\author[0000-0002-7633-2883]{David J. Lagattuta}
\affiliation{Univ Lyon, Univ Lyon1, Ens de Lyon, CNRS, Centre de Recherche Astrophysique de Lyon UMR5574, F-69230, Saint-Genis-Laval, France}
\affiliation{Centre for Extragalactic Astronomy, Durham University, South Road, Durham DH1 3LE, U.K.}
\affiliation{Institute for Computational Cosmology, Durham University, South Road, Durham DH1 3LE, U.K.}

\author{Marceau Limousin}
\affiliation{Aix Marseille Univ, CNRS, CNES, LAM, Marseille, France}

\author{Anja von der Linden}
\affiliation{Department of Physics and Astronomy, Stony Brook University, Stony Brook, NY 11794, USA}

\author[0000-0001-8840-2538]{Nora B. Linzer}
\affiliation{Cosmic Dawn Center (DAWN)}
\affiliation{Division of Physics, Mathematics, and Astronomy, California Institute of Technology, Pasadena, CA 91125, USA}

\author{Adrian Lopez}
\affiliation{California Institute of Technology, 1200 East California Boulevard, Pasadena, CA 91125, USA}
\affiliation{Cosmic Dawn Center (DAWN)}

\author[0000-0002-4872-2294]{Georgios E. Magdis}
\affiliation{Cosmic Dawn Center (DAWN)}
\affiliation{DTU-Space, Technical University of Denmark, Elektrovej 327, DK-2800 Kgs. Lyngby}
\affiliation{Niels Bohr Institute, University of Copenhagen, Lyngbyvej 2, Copenhagen \O~2100}

\author[0000-0002-6085-3780]{Richard Massey}
\affiliation{Centre for Extragalactic Astronomy, Durham University, South Road, Durham DH1 3LE, U.K.}

\author{Daniel C. Masters}
\affiliation{Jet Propulsion Laboratory, California Institute of Technology, Pasadena, CA 91109, USA}

\author{Matteo Maturi}
\affiliation{Zentrum f\"ur Astronomie, Universit\"at Heidelberg, Philosophenweg 12, D-69120 Heidelberg, Germany}

\author[0000-0001-5807-7893]{Curtis McCully}
\affiliation{Las Cumbres Observatory, 6740 Cortona Drive, Suite 102, Goleta, CA 93117-5575, USA}
\affiliation{Department of Physics, University of California, Santa Barbara, CA 93106-9530, USA}

\author[0000-0003-3255-3139]{Sean L. McGee}
\affiliation{School of Physics and Astronomy, University of Birmingham, Birmingham B15 2TT, United Kingdom}

\author[0000-0003-1225-7084]{Massimo Meneghetti}
\affiliation{INAF - Osservatorio di Astrofisica e Scienza dello Spazio, Via Gobetti 93/3, 40129 Bologna, Italy}

\author{Bahram Mobasher}
\affiliation{Physics and Astronomy Department, University of California, Riverside, CA 92521, USA}

\author[0000-0003-3030-2360]{Leonidas A.\ Moustakas}
\affiliation{Jet Propulsion Laboratory, California Institute of Technology, Pasadena, CA 91109, USA}

\author[0000-0001-7089-7325]{Eric J. Murphy}
\affiliation{National Radio Astronomy Observatory, Charlottesville, VA 22903, USA}

\author{Priyamvada Natarajan}
\affiliation{Department of Astronomy, Yale University, New Haven CT 06511, USA}

\author[0000-0002-2618-5790]{Mark Neyrinck}
\affiliation{Ikerbasque, Basque Foundation for Science, E-48011 Bilbao, Spain}
\affiliation{Department of Theoretical Physics, University of the Basque Country UPV/EHU, E-48080 Bilbao, Spain;}

\author{Kyle O'Connor}
\affiliation{University of South Carolina, Department of Physics \& Astronomy, Columbia, SC, USA}

\author{Masamune Oguri}
\affiliation{Research Center for the Early Universe, University of Tokyo, Tokyo 113-0033, Japan}
\affiliation{Department of Physics, University of Tokyo, Tokyo 113-0033, Japan}
\affiliation{Kavli Institute for the Physics and Mathematics of the Universe (Kavli IPMU, WPI), University of Tokyo, Chiba 277-8582, Japan}

\author[0000-0002-6015-8614]{Amanda Pagul}
\affiliation{Department of Physics and Astronomy, University of California, Riverside 92521, USA}

\author{Jason Rhodes}
\affiliation{Jet Propulsion Laboratory, California Institute of Technology, Pasadena, CA 91109, USA}
\affiliation{Kavli Institute for the Physics and Mathematics of the Universe (Kavli IPMU, WPI), UTIAS, The University of Tokyo, Chiba 277-8583, Japan}

\author[0000-0003-0427-8387]{R. Michael Rich}
\affiliation{Department of  Physics and Astronomy, UCLA, PAB 430 Portola Plaza, Box 951547, 90095-1547}

\author[0000-0002-0086-0524]{Andrew Robertson}
\affiliation{Institute for Computational Cosmology, Durham University, South Road, Durham DH1 3LE, U.K.}

\author[0000-0003-0302-0325]{Mauro Sereno}
\affiliation{INAF - Osservatorio di Astrofisica e Scienza dello Spazio di Bologna, via Piero Gobetti 93/3, I-40129 Bologna, Italy}
\affiliation{INFN, Sezione di Bologna, viale Berti Pichat 6/2, 40127 Bologna, Italy}

\author{Huanyuan Shan}
\affiliation{Shanghai Astronomical Observatory (SHAO), Nandan Road 80, Shanghai, China}

\author[0000-0003-4494-8277]{Graham P. Smith}
\affiliation{School of Physics and Astronomy, University of Birmingham, Edgbaston, B15 2TT, England}

\author{Albert Sneppen}
\affiliation{Niels Bohr Institute, University of Copenhagen, Lyngbyvej 2, Copenhagen \O~2100}
\affiliation{Cosmic Dawn Center (DAWN)}

\author{Gordon K. Squires}
\affiliation{IPAC, California Institute of Technology, 1200 E. California Blvd., Pasadena, CA, 91125}

\author{Sut-Ieng Tam}
\affiliation{Institute for Computational Cosmology, Durham University, South Road, Durham DH1 3LE, U.K.}

\author[0000-0003-4219-3683]{C\'eline Tchernin}
\affiliation{Center for Astronomy, Institute for Theoretical Astrophysics, Heidelberg University, Philosophenweg 12, 69120, Heidelberg, Germany}

\author[0000-0003-3631-7176]{Sune Toft}
\affiliation{Cosmic Dawn Center (DAWN)}
\affiliation{Niels Bohr Institute, University of Copenhagen, Lyngbyvej 2, Copenhagen \O~2100}

\author[0000-0002-7196-4822]{Keiichi Umetsu}
\affiliation{Academia Sinica Institute of Astronomy and Astrophysics (ASIAA), No. 1, Section 4, Roosevelt Road, Taipei 10617, Taiwan}

\author[0000-0003-1614-196X]{John R. Weaver}
\affiliation{Cosmic Dawn Center (DAWN)}
\affiliation{Niels Bohr Institute, University of Copenhagen, Lyngbyvej 2, Copenhagen \O~2100}

\author[0000-0002-0587-1660]{R. J. van Weeren}
\affiliation{Leiden Observatory, Leiden University, PO Box 9513, 2300 RA Leiden, The Netherlands}

\author[0000-0002-6039-8706]{Liliya L. R. Williams}
\affiliation{School of Physics and Astronomy, University of Minnesota, 116 Church St SE, Minneapolis MN 55455, USA}

\author[0000-0001-6352-9735]{Tom J. Wilson}
\affiliation{Space Telescope Science Institute, 3700 San Martin Dr., Baltimore, MD 21218, USA}

\author[0000-0003-1710-9339]{Lin Yan}
\affiliation{The Caltech Optical Observatories, California Institute of Technology, Pasadena, CA 91125, USA}

\author[0000-0002-0350-4488]{Adi Zitrin}
\affiliation{Physics Department, Ben-Gurion University of the Negev, P.O. Box 653, Be’er-Sheva 84105, Israel}





\begin{abstract}
The Beyond Ultra-deep Frontier Fields and Legacy Observations (BUFFALO) is a 101 orbit + 101 parallel Cycle 25 Hubble Space Telescope Treasury program taking data from 2018-2020.  BUFFALO will expand existing coverage of the Hubble Frontier Fields (HFF) in WFC3/IR F105W, F125W, and F160W and ACS/WFC F606W and F814W around each of the six HFF clusters and flanking fields.  This additional area has not been observed by HST but is already covered by deep multi-wavelength datasets, including Spitzer and Chandra.  As with the original HFF program, BUFFALO is designed to take advantage of gravitational lensing from massive clusters to simultaneously find high-redshift galaxies which would otherwise lie below HST detection limits and model foreground clusters to study properties of dark matter and galaxy assembly.  The expanded area will provide a first opportunity to study both cosmic variance at high redshift and galaxy assembly in the outskirts of the large HFF clusters.  Five additional orbits are reserved for transient followup.  BUFFALO data including mosaics, value-added catalogs and cluster mass distribution models will be released via MAST on a regular basis, as the observations and analysis are completed for the six individual clusters.
\end{abstract}

\section{Introduction} 
\label{sec:intro}

The Beyond Ultra-deep Frontier Fields and Legacy Observations (BUFFALO) program is a Hubble Treasury program expanding the Hubble Frontier Fields galaxy cluster to adjacent areas that already benefit from ultra-deep \emph{Spitzer} and multi-wavelength coverage.  The program consists of 96 orbits + 96 parallel orbits in the Frontier Fields and also includes five orbits of planned supernova followup, based upon expected rates.

The Frontier Fields program \citep{Lotz2017} arose out of the realization that the same dataset could be used to attack two of the most important questions in astronomy, even though they were seemingly different topics. Gravitational lensing from massive, foreground clusters allows the \emph{Hubble Space Telescope} (\emph{HST}) to detect galaxies which would otherwise be too faint.  At the same time, these observations allow us to model the dark matter distribution in the foreground clusters, which provides some of the leading constraints on the properties of dark matter.  

BUFFALO will build upon the existing Frontier Fields programs by significantly broadening the area observed by \emph{Hubble} around each of the six Frontier Fields clusters.  BUFFALO will observe four times the area of the existing Frontier Fields in a tiling centered around the central region of each cluster field (and, in parallel, flanking field).  BUFFALO will observe these new regions in five \emph{HST} filters, including most of the filters used for the original Frontier Fields program.  Observations over this broader region will provide significant improvements in both our understanding of dark matter and of high-redshift galaxies.

Both of these studies benefit strongly from the presence of multi-wavelength data that has been added to \emph{HST} by the \emph{Spitzer Space Telescope} \citep{Werner2004}, the \emph{Chandra X-ray Observatory} \citep{Weisskopf2000}, the \emph{XMM-Newton X-ray Observatory} \citep{Jansen2001}, and other ground-based observatories.  BUFFALO is designed to take advantage of this by expanding Frontier Fields coverage to fields in which these other data already exist. 

The design benefits additionally from previous \emph{HST} studies of high-redshift galaxies \citep{Oesch2010,McLure2010,Finkelstein2015,Bouwens2015,Bouwens2016}.  Their measurements of the high-redshift luminosity function enable BUFFALO to choose a depth in five optical and near-infrared pass-bands with the Advanced Camera for Survey (ACS) and the Wide Field Camera 3 (WFC3), that is optimized to select the most luminous galaxies at $z>6$ and to study their physical properties such as stellar masses by fully exploiting the pre-existing, deep \emph{Spitzer}/IRAC data around the Frontier Fields. BUFFALO is thus the logical extension of the Hubble Frontier Fields program, HFF \citep{Lotz2017} 
as well as previous \emph{HST+Spitzer} extragalactic legacy surveys such as
GOODS \citep{Giavalisco2004}, HUDF \citep{Beckwith2006,Ellis2013,Illingworth2013,Labbe2015}, CANDELS \citep{Grogin2011,Koekemoer2011}, S-CANDELS \citep{Ashby2015}, COSMOS \citep{Scoville2007}, SMUVS \citep{Ashby2018}, BORG/HIPPIES \citep{Trenti2011,Yan2011}, CLASH \citep{Postman2012}, SURFS UP \citep{Bradac2014}, or RELICS \citep{Coe2019}. 

The BUFFALO observations and filter choices are summarized in \S~\ref{sec:observations}.  The BUFFALO collaboration includes groups which use a variety of techniques for mass modeling for individual clusters, as discussed in \S~\ref{sec:lensingmm}.  In \S~\ref{sec:highz}, the key science goals for high-redshift galaxy evolution are described.  The improvements in our understanding of structure evolution and dark matter physics from expanding the \emph{Hubble} footprint to a wider area around these clusters are discussed in \S~\ref{sec:structure}.  As discussed in \S~\ref{sec:supernovae}, BUFFALO is also expected to observe a variety of transient objects, possibly including lensed supernovae.  Planned data products are described in \S~\ref{sec:dataproducts}.  The survey program and key results are briefly summarized in \S~\ref{sec:summary}.

Analysis presented here uses the AB magnitude system \citep{Oke1974,Gunn1983} and a flat $\Lambda$CDM cosmology with $(h, \Omega_m, \Omega_\Lambda) = (0.674 ,0.315, 0.685)$ \citep{Planck2018} throughout.

\section{Observations}
\label{sec:observations}

\begin{figure}
\includegraphics[width=0.47\textwidth]{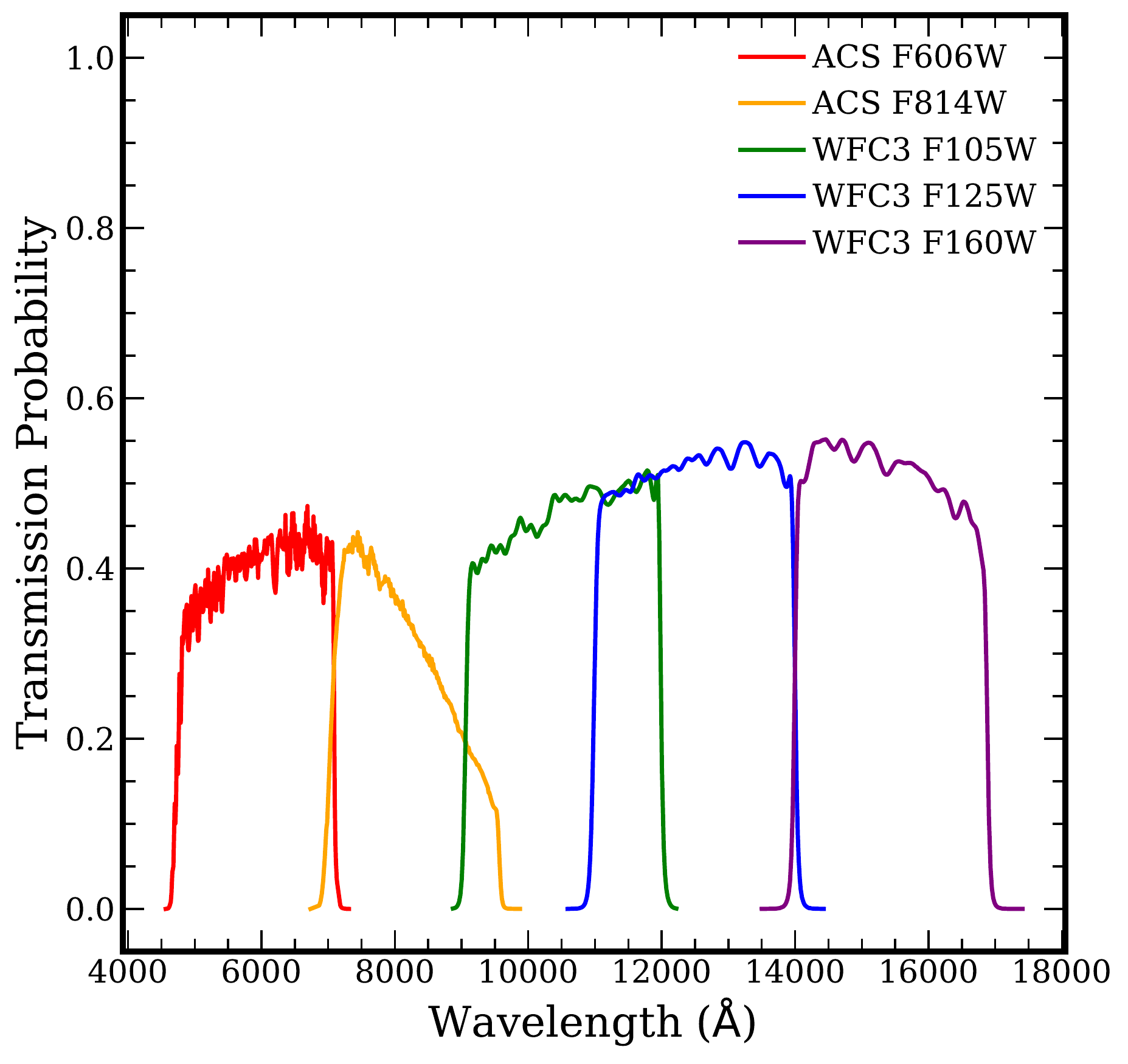}
\caption{\emph{HST} filter profiles for BUFFALO observations using ACS and WFC3.  BUFFALO uses many of the same filters as the original Hubble Frontier Fields program, with 2/3 of an orbit depth in WFC3/F105W, WFC3/F125W, WFC3/F160W, and ACS/F606W and 4/3 orbit depth in ACS/F814W.}
\label{fig:filters}
\end{figure}

\begin{figure*}
\includegraphics[width=\textwidth]{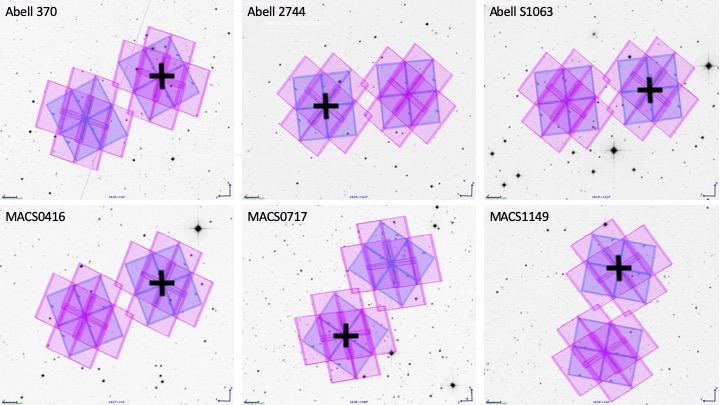}
\caption{BUFFALO coverage map for the six Frontier Field clusters, produced by running the HST Astronomer's Proposal Tool (APT) on the full set of observations from this program (ID 15117).  Each BUFFALO pointing, chosen to overlap the regions with deep Channel 1 and Channel 2 data from \emph{Spitzer}/IRAC, is covered by 2/3 orbit depth WFC3 exposures in F105W, F125W and F160W (blue) and by ACS observations taken in parallel, at 2/3 orbit depth in F606W and 4/3 orbit depth in F814W (pink). The ACS imaging fully covers the WFC3-IR area and goes somewhat beyond it due to the larger field of view of ACS. The new BUFFALO tiling is centered at the central region of each cluster field, indicated with a black "plus" sign in each case, with parallel observations similarly expanding the existing flanking fields.}
\label{fig:buffalomap}
\end{figure*}

 The observing strategy for BUFFALO is designed to increase the area observed by \emph{HST} around each of the six Frontier Fields (HFF) clusters (Table \ref{tab:clusters}), with both ACS/WFC and WFC3/IR cameras, and in many of the same filter sets used for the HFF program (Fig. \ref{fig:filters}). Specifically, each of the six HFF clusters is observed for a total of 16 orbits, divided into two orients (8 orbits each) differing by 180 degrees (approximately six months apart), thereby providing ACS/WFC and WFC3/IR coverage of both the main cluster field and the parallel field. In addition, each of the 8-orbit orients is divided into two epochs (at 4 orbits per epoch), separated by about 30 days to enable searches for Type Ia supernovae in both the main cluster and the parallel field. Each 4-orbit epoch consists of a 2$\times$2 mosaic across the field, with a spacing designed to maximize the area covered by the WFC3/IR detector, thereby covering four times the area of the original WFC3/IR imaging, for both the main cluster and the parallel field, in all cases. The corresponding ACS observations overlap each other by about half the field of view of ACS, with the resulting mosaic covering three times the area of the original ACS imaging, again for both the main cluster and the parallel field (Fig.~\ref{fig:buffalomap}).

\begin{deluxetable*}{lccc}[t]
\tablecaption{\label{tab:clusters}
	BUFFALO/Hubble Frontier Fields Clusters and Flanking Fields}
\tablehead{%
Field	        	& Center (J2000)	& Flanking Field Center (J2000)}
\startdata
Abell 370           & 02:39:52.9 -01:34:36.5 & 02:40:13.4 −01:37:32.8 \\
					&		    &   	   	        	\\
MACS J0717.5+3745	& 07:17:34.0 +37:44:49.0 & 07:17:17.0 +37:49:47.3	\\
					&	    	&   	   				        	\\
MACS J0416.1-2403	& 04:16:08.9 -24:04:28.7 & 04:16:33.1 -24:06:48.7	\\
					&	    	&   	   				        	\\
Abell S1063		    & 22:48:44.4 -44:31:48.5 & 22:49:17.7 -44:32:43.8	\\
					&	    	&	       					        \\
Abell 2744		    & 00:14:21.2 -30:23:50.1 & 00:13:53.6 -30:22:54.3	\\
					&	    	&   	   			        	\\
MACS J1149.5+2223	& 11:49:36.3 +22:23:58.1 & 11:49:40.5 +22:18:02.3	\\
\enddata
\tablecomments{Please see \citet{Lotz2017} for additional information about the original Hubble Frontier Fields pointings and cluster properties.}
\end{deluxetable*}

Once all 16 orbits are obtained for each cluster, the resulting observations comprise a 2$\times$2 mosaic of WFC3/IR data in F105W, F125W and F160W, each at approximately 2/3 orbit depth, utilizing a 4-point dither to mitigate detector defects and also provide sub-pixel sampling. The corresponding ACS/WFC observations consist of F606W at 2/3 orbit depth, and F814W at 4/3 orbit depth, again utilizing a 4-point dither to mitigate detector defects, provide sub-pixel sampling and step across the gap between the two chips in ACS.
For each of the six clusters, the final dataset therefore consists of 160 exposures in total, at 4 exposures per pointing, over 8 different pointings (4 on the main cluster and 4 on the parallel field), for each of the five ACS/WFC3 filters used in the program.

As each new set of observations were obtained, they were downloaded and processed with a series of steps that go beyond the standard STScI pipeline processing. In particular, these additional steps included refining the astrometry for all exposures to accuracies better than milliarcsecond-level, as well as improved background level estimates and refined cosmic ray rejection, and improved treatment of time-variable sky and persistence removal for the WFC3/IR detectors. These improvements all followed similar procedures to those described in \citet{Koekemoer2011}, where more details on these steps are presented.  Exposures in the F105W band are affected by time-varying background signal due to variable atmospheric emission during the exposure. As a result, the background of an exposure is  higher and the final depth of the images can be impacted. For its characteristics, diffuse and extended, this effect can impact the detection of intracluster light (see \S~\ref{subsec:icl}) in this band. The BUFFALO image processing corrects for the time variable background by either subtracting a constant background for each readout or excluding the affected frames (see \citealt{Koekemoer2013} for more details on the processing).  A preliminary inspection of the data shows no difference between the intracluster light maps for the F105W and the other IR bands.

The final products prepared from these procedures included full-depth combined mosaics for each filter, as well as individual exposures necessary for lensing models, all astrometrically aligned and distortion-rectified onto a common astrometric pixel grid, which was tied directly to the GAIA-DR2 absolute astrometric frame using catalogs provided by M. Nonino (priv. comm). The final mosaics produced from this process also included all other previous HST data obtained on these fields in these filters, and therefore are the deepest \emph{HST} mosaics available for these fields.
 
 \begin{deluxetable*}{llllll}[t]
\tablecaption{\label{tab:obs-schedule}
	BUFFALO {\it HST} WFC3 and ACS Observing Schedule}
\tablehead{%
Field	        	& Visit Numbers	& Main Cluster	& Parallel Field    & Epoch 1   & Epoch 2       }
\startdata
Abell 370           & 6A-6H		& WFC3	& ACS	& 2018 Jul 21-22		& 2018 Aug 21			\\
					& 6I-6P		& ACS	& WFC3	& 2018 Dec 19			& 2019 Jan 7			\\
					&		    &   	&   	&			        	&			        	\\
MACS J0717.5+3745	& 3I-3P		& ACS	& WFC3	& 2018 Oct 2-3			& 2018 Nov 22			\\
					& 3A-3H		& WFC3	& ACS	& 2019 Feb 18			& 2019 Apr 1			\\
					&	    	&   	&   	&		        		&			        	\\
MACS J0416.1-2403	& 2I-2P		& ACS	& WFC3	& 2019 Jan 7			& 2019 Feb 7			\\
					& 2A-2H		& WFC3	& ACS	& 2019 Aug 3		& 2019 Sep 6	\\
					&	    	&   	&   	&		        		&			        	\\
Abell S1063		    & 5A-5H		& WFC3	& ACS	& 2019 Apr 20		& 2019 May 29	\\
					& 5I-5P		& ACS	& WFC3	& 2019 Oct 			& 2019 Nov 			\\
					&	    	&	    &   	&			        	&				        \\
Abell 2744		    & 1I-1P		& ACS	& WFC3	& 2019 May 15		& 2019 Jul 3	\\
					& 1A-1H		& WFC3	& ACS	& 2019 Oct - Nov	& 2019 Nov - Dec	\\
					&	    	&   	&   	&			        	&			        	\\
MACS J1149.5+2223	& 4A-4H		& WFC3	& ACS	& 2019 Dec	    		& 2020 Jan		    	\\
					& 4I-4P		& ACS	& WFC3	& 2019 Apr		    	& 2020 May		    	\\
\enddata
\end{deluxetable*}

\subsection{Analysis Tools and Methods}
 
In order to allow analysis of the intrinsic properties of background lensed galaxies, the BUFFALO survey follows the HFF philosophy of releasing high-level data products through the MAST archive for the scientific community. The BUFFALO team will produce several strong, weak, and joint strong$+$weak lensing models from independent teams using different techniques described further in \S~\ref{sec:lensingmm}. BUFFALO will then deliver mass, magnification, deflection, and associated error maps for different redshifts provided by each independent modeling team of the collaboration, as well as weak-lensing catalogues.  An online magnification calculator will also be available for fast magnification estimates and errors within the modeled field-of-views for each modeling pipeline.

\section{Lensing Mass Modeling}
\label{sec:lensingmm}

The BUFFALO survey extends the \textit{Hubble Space Telescope} coverage of the 6 \textit{Hubble Frontier Fields} \citep[HFF,][]{Lotz2017} clusters by a factor of about 4, which will expand the measurements of the overall mass profile and substructure characteristics of both the dark and luminous components of galaxy clusters up to  $\sim 3/4 \times\mathrm{R_{\rm{vir}}}$. Recent studies have shown that the shear induced by distant and massive substructures can extend to the cluster core \citep{Acebron2017, Mahler2018, Lagattuta2019}, and therefore including them in the overall mass modeling can affect the central component of the model and remove some potential biases. \citet{Mahler2018} showed that structures in the outskirts of \textit{Abell\,2744} found with a weak lensing analysis \citep{Jauzac2016} impact the mass measurements in the cluster core, which in turn affects magnification estimates of high redshift sources. \citet{Lagattuta2019} studied Abell\,370 by extending the \textit{Multi-Unit Spectroscopic Explorer} (MUSE) mosaic around Abell\,370 together with deeper imaging data from the HFF program. Their best-fit strong lensing model needed a significantly large external shear term that is thought to come from structures along the line-of-sight and at larger projected radii in the lens plane. 

Moreover, \citet{Acebron2017} analyzed the impact of distant and massive substructures on the overall mass and density profiles reconstruction with \emph{HFF}-like simulated clusters \citep{Meneghetti2017}. They found that the density profile in the outskirts of the cluster ($\geq 600 kpc$) was underestimated by up to $\sim30\%$. Including the substructures in the modeling helps to better constrain the mass distribution, with an improvement between $\sim5\%$ to $\sim20\%$ in the most distant regions of the cluster. Further, the bias introduced by these unmodeled structures also affected the retrieval of cosmological parameters with the strong lensing cosmography technique.

Mapping these structures is essential not only for improving the accuracy of the mass models, but also to gain a better understanding of the physical properties governing galaxy clusters.  Indeed, as the densest regions of the Universe and the sites of constant growth due to their location at the nodes of the cosmic web, galaxy clusters represent a privileged laboratory for testing the properties of dark matter and for improving our global picture of structure evolution. These recent studies have thus further proven the need for extended data coverage around galaxy clusters.

With its extended spatial coverage of the HFF clusters, BUFFALO will provide unprecedented weak-lensing measurements thanks to \emph{HST} high-resolution imagery, allowing for precise measurement of the shapes of weakly-lensed galaxies. The combination of BUFFALO weak-lensing and HFF strong-lensing analyses will provide the most precise mass measurements ever obtained for those clusters. BUFFALO will detect structures that can only be significantly detected with \textit{HST} weak-lensing. While residing in the cluster outskirts, they still introduce a significant bias in the mass measurements as explained earlier, and thus on the magnification estimates that are crucial for high redshift studies. 

Following the success of a similar program with the original HFF observations, the lensing profile in BUFFALO will be modeled by several independent teams, with each result released for public use (\S~\ref{tab:rel-schedule}).  The aim of the BUFFALO mass modeling challenge is to bring a better understanding of the behaviour of these clusters as well as on the different systematic errors arising from the different techniques or assumptions used \citep{Meneghetti2017}. 
The algorithms participating in the modeling challenge are briefly described below: \\ 
\begin{itemize}
\item {\tt glafic} \citep{Oguri2010} performs mass modeling adopting a parametric approach. Halo components are remodeled by an Navarro-Frenk-White (NFW) profile, and cluster member galaxies are modeled by a pseudo-Jaffe profile. External perturbations are added for better modeling. A detailed description of the mass modeling of the \textit{HFF} clusters with {\tt glafic} is given in \citet{Kawamata2016} and \citet{Kawamata2018}, indicating that positions of multiple images are reproduced with a typical accuracy of $\sim 0.4''$.
    
\item \texttt{Grale} \citep{Liesenborgs2006, Liesenborgs2009} is a flexible, free-form method, based on a genetic algorithm that uses an adaptive grid to iteratively refine the mass model. As input it uses only the information about the lensed images, and nothing about the cluster's visible mass. This last feature sets \texttt{Grale} apart from most other lens mass reconstruction techniques, and gives it the ability to test how well mass follows light on both large and small scales within galaxy clusters. \texttt{Grale}'s description, software and installation instructions are available online at {\url{http://research.edm.uhasselt.be/~jori/grale}}.

\item {\textsc{Lenstool} \citep{Jullo2007, Jullo2009} utilizes a Bayesian Markov Chain Monte-Carlo sampler to optimize the model parameters using the positions and spectroscopic redshifts (as well as magnitudes, shapes and multiplicity if specified) of the multiply imaged systems. The overall matter distribution of clusters is decomposed into smooth large scale components and the small-scale halos associated with the locations of cluster galaxies. 

A new version of \textsc{Lenstool} (Niemiec et al., in prep.) allows the combination a parametric strong lensing modeling of the cluster's core, where the multiple images appear, with a more flexible non-parametric weak lensing modeling at larger radii, the latter leading to the detection and characterization of substructures in the outskirts of galaxy clusters \citep{Jauzac2016,Jauzac2018}.}
    
\item The \textsc{Light-Traces-Mass} \citep{Broadhurst2005, Zitrin2009} methodology is based on the assumption that the underlying dark matter distribution in the cluster is traced by the distribution of the luminous component, i.e., the cluster galaxies and their luminosities. Only a small number of free parameters is needed to generate a mass model where the position and source redshift of multiple images are used as constraints. \textsc{LTM} has proven to be a powerful method for identifying new multiple images as well as for constraining the cluster mass distribution \citep{Merten2011, Zitrin2015, Frye2019}. This pipeline also allows the incorporation of other constraints including weak-lensing shear measurements, time delays, or relative magnifications.
\item The \textsc{WSLAP+} \citep{Diego2007,Diego2016} method falls in the category of hybrid methods. The mass distribution is decomposed into two components: (1) a grid component (free-form method) which accounts for the diffuse mass distribution, and (2) a compact component (parametric method) which accounts for the mass associated with the member galaxies. The method combines strong-lensing arc positions with weak-lensing measurements when available. The solution is found by minimizing the quadratic form of a system of linear equations. The role of the regularization is adopted by the number of iterations in the minimization code, although if resolved lensed systems are present (i.e, arcs with multiple identifiable knots) the solutions derived are very robust and weakly dependent on the number of iterations.  Details can be found in \citet{Diego2007,Diego2016} and references therein. 

\item {The \texttt{CLUMI} (CLUster lensing Mass Inversion) code \citep{Umetsu2013} combines wide-field weak-lensing (shear and magnification) constraints for reconstructing binned surface mass density profiles,  $\Sigma(R)$. This code has been used for the CLASH weak-lensing studies \cite[e.g.,][]{Umetsu2014}. It can also include central strong-lensing constraints in the form of the enclosed projected total mass, as done in \citet{Umetsu2016}. 

The \texttt{CLUMI}-2D code \citep{Umetsu2018} is a generalization of \texttt{CLUMI} into a 2D description of the pixelized mass distribution. It combines the spatial 2D shear pattern (g1, g2) with azimuthally-averaged magnification-bias measurements, which impose a set of azimuthally integrated constraints on the $\Sigma(x, y)$ field, thus effectively breaking the mass-sheet degeneracy. It is designed for an unbiased reconstruction of both mass morphology (e.g., halo ellipticity) and the radial mass profile of the projected cluster matter distribution.}
\end{itemize}

\section{Impact on Studies of the High-redshift Universe}
\label{sec:highz}

Massive galaxies at $z\sim8-10$ are in the midst of a critical transition between initial collapse and subsequent evolution \citep{Steinhardt2014a,Bouwens2015,Steinhardt2016,Mashian2016}. Galaxies are thought to form via hierarchical merging, building from primordial dark matter fluctuations into massive proto-galaxies \citep{millennium,Illustris}.  Indeed, at $z<6$, the stellar mass function (SMF) is well approximated by a Schechter function with a clear exponential cutoff above a characteristic mass.  

However, if early galactic assembly is dominated by simple baryonic cooling onto dark matter halos, at very high redshifts the SMF should be a Press-Schechter-like power law with no exponential cutoff.  Similarly, studies of star forming galaxies at $z<6$ \citep{Noeske2007,Daddi2007,Lee2013,Speagle2014} find a relatively tight relationship between the stellar mass and star formation rate (SFR) of star-forming galaxies (the so-called star forming ``main sequence''), but according to the hierarchical growth paradigm, increased merger rates should result in a much weaker correlation of these two quantities at high redshift.  Observing this transition would pinpoint the end of the initial growth phase, providing strong constraints for models of early structure formation and the nature and causes of reionization.  

This transition is currently not well constrained at $z\gtrsim6$ due to the small number of galaxies with the necessary SFR \emph{and} stellar mass estimates required, which at present only the combination of HST/WFC3 and Spitzer-IRAC can provide. However, recent results using HST photometry \citep{Bouwens2015,Bouwens2016,Davidzon2017} indicate that the galaxy luminosity function flattens between $z \sim 7$ (where there is likely an exponential cutoff) and $z \sim 10$ (where no cutoff is observed), consistent with the epoch around $z \sim 8-10$ is the critical transition window.  

\begin{figure}
\includegraphics[width=0.47\textwidth]{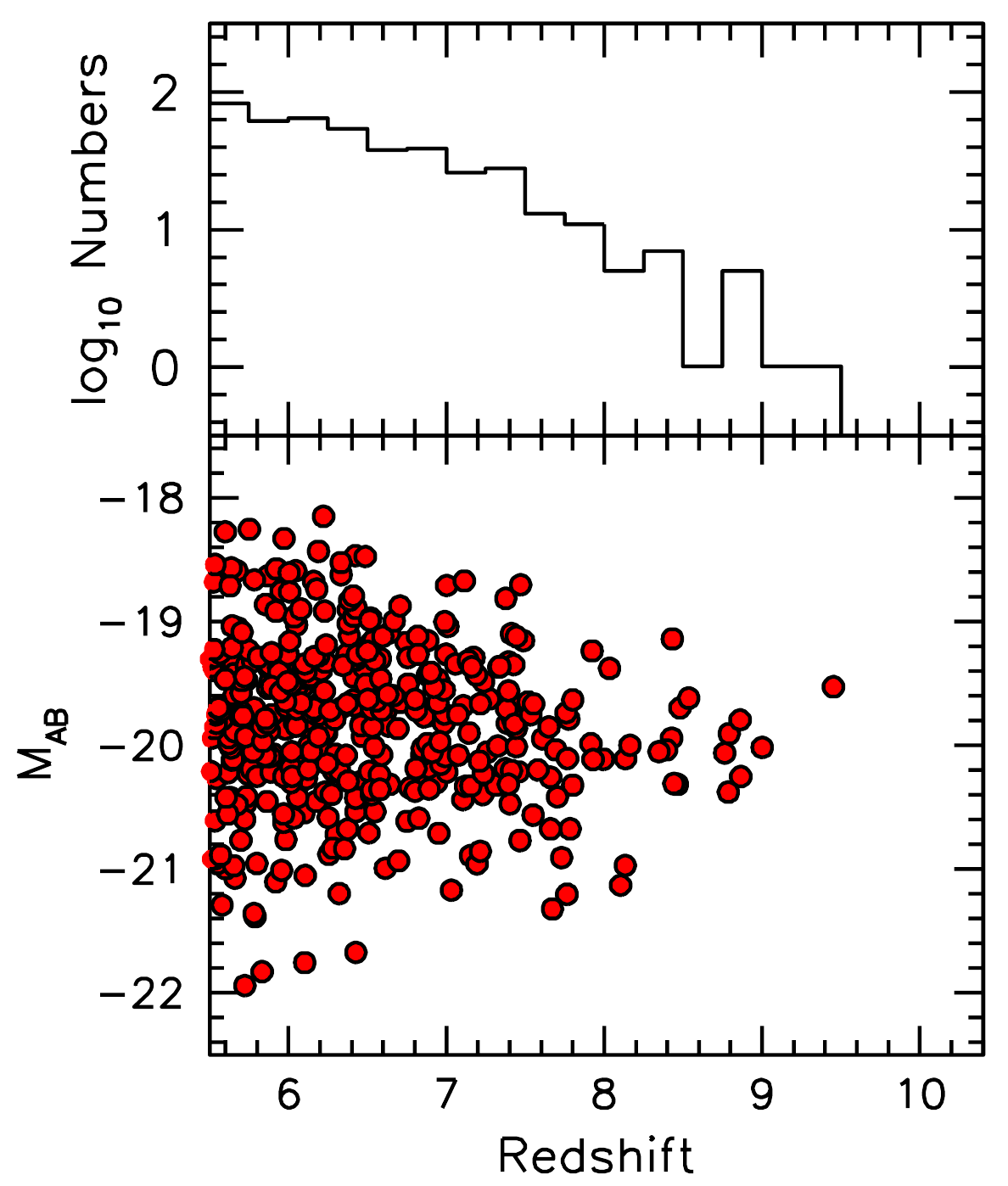}
\caption{Simulated high-redshift, luminous galaxies (bottom) and overall distribution (top) in a random selection corresponding to the BUFFALO clusters and flanking fields.  It is likely that the highest-redshift BUFFALO galaxy will lie at $z\sim9-10$.}
\label{fig:galcounts}
\end{figure}

Uncertainties at this epoch are currently dominated by a combination of cosmic variance, insufficient area in current sightlines, and lack of ultra-deep Spitzer data needed to confirm luminous $z>8$ systems.  To illustrate the point, all published $z\sim 9-10$ CANDELS galaxies are in just two of the five possible fields where they could be discovered \citep{Oesch2014,RobertsBorsani2016}.  The BUFFALO survey is optimized to efficiently search for these early, high-mass galaxies over the remaining areas where mass estimation will be possible (Fig. \ref{fig:galcounts}).
   
In addition to the expected change in the shape of the galaxy mass function, recent studies using HST and Spitzer observations \citep{Lee2013,Stark2013,Steinhardt2014a,Bouwens2015,Bouwens2016,Oesch2016} find a substantial population of galaxies that are seemingly \emph{too massive, too early} to have formed through standard hierarchical merging \citep{Steinhardt2016,Behroozi2018b}.  However, these massive galaxies are also expected to be the most strongly biased and thus measurements are most affected by cosmic variance, and the factor of $\sim$2 improvement in their measured number density provided by BUFFALO will either relieve or significantly increase this tension.

BUFFALO is designed to optimally reduce the current uncertainty by imaging sightlines with existing, sufficient ultra-deep Spitzer data.  Joint HST and Spitzer observations have proven to be essential for the discovery and characterization of the highest-redshift galaxies and have recently resulted in the first rest-frame optical detection and stellar mass measurements of individual galaxies at z$\sim$8 and even z$\sim$10 \citep{Labbe2013,Oesch2013}.

Because early, massive galaxies are very highly biased \citep{Moster2011}, the limiting factor in our understanding of this transition period is a combination of cosmic variance and area coverage.  Of the three $z = 10$ galaxy candidates in $\sim$800~arcmin$^{2}$ of deep HST imaging over the five CANDELS sightlines, two lie in a single 4~arcmin$^{2}$ region \citep{Oesch2014}, and the $z>7$ candidates in the HFF fields are highly clustered. The addition of BUFFALO data will improve the measurement of the space density of L$>$L$_*$ galaxies by doubling the number of sightlines with sufficient area, filter coverage, and Spitzer data to reliably detect galaxies at z$\mathbf{\sim}$8-10. 

Spitzer data are essential for $z>8$ galaxy studies (cf. \citet{Bradac2014}, Fig. 3) because they constrain the age and mass of high-redshift galaxies and can be used to effectively remove $z<3$ contaminants. Spitzer has already invested $\sim$1500h (2 months) imaging the HFF areas to 50-75h depth, but HST WFC3-IR data exists over only $\lesssim$10\% of this deep Spitzer coverage.  The BUFFALO pointings were chosen to overlap with the existing Spitzer HFF coverage as efficiently as possible in regions central enough that there is still likely to be non-negligible magnification from weak lensing.  The BUFFALO coverage areas also expands WFC3 coverage to an area which is well matched with the \emph{JWST} NIRSPEC field of view.


BUFFALO will complete the WFC3-IR and ACS coverage of areas with Spitzer-IRAC deep enough to study the high-$z$ universe over a large area, resulting in the best study of this key period currently possible.  Critically, the large area and additional sightlines will both mitigate the effect of cosmic variance due to the strong clustering of $z>8$ objects \citep{Trenti2012,Oesch2013,Zheng2014,Laporte2014} and directly constrain the galaxy bias with respect to the dark matter \citep{Adelberger1998,Robertson2010}.

Presently, Spitzer-IRAC data at depths greater than 50h per pixel in both 3.6$\mu$m and 4.5$\mu$m bands are the \emph{only} way to measure the stellar masses of galaxies at $z\sim 8-10$ and firmly establish that they are indeed high-redshift (e.g. Fig. 5), while HST data are crucial to identifying candidate $z\sim 8-10$ galaxies through their photometric dropout.  This powerful combination of data currently exists over $<740~ \mathrm{arcmin}^{2}$ of sky, yielding $\sim$20 joint detections of $z\sim 8$ galaxies and $\sim$4 at $z\sim10$.  The strong clustering of high-redshift galaxies has been definitively observed, with the majority of $z\sim10$ galaxies found in one of five CANDELS fields \citep{Bouwens2015} and one of the Frontier Fields known to contain an over-density of $z\sim8$ galaxies in the region mapped by BUFFALO \citep{Zheng2014,Ishigaki2015}.

\begin{figure}
\includegraphics[width=0.47\textwidth]{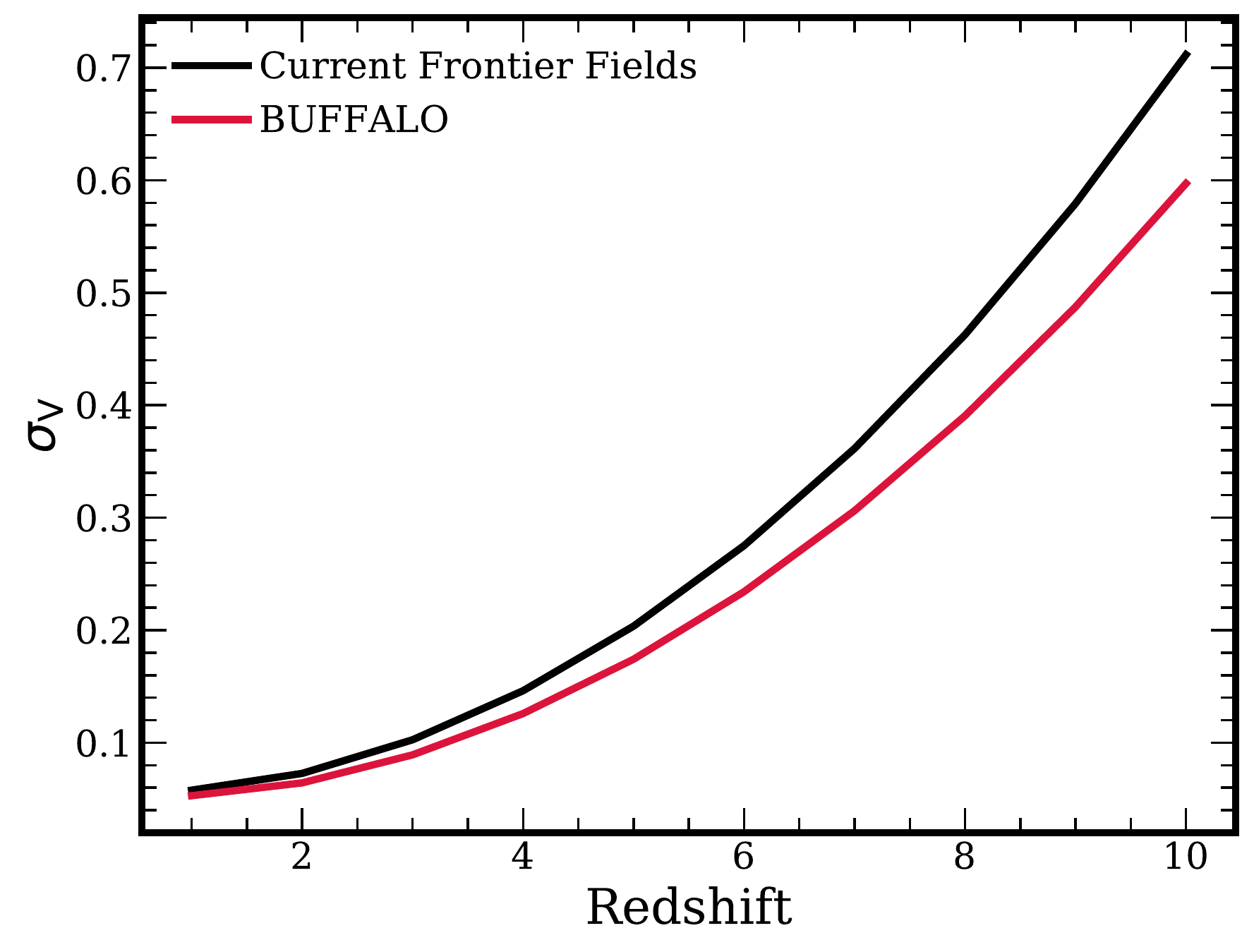}
\caption{Predicted cosmic variance $\sigma_v$ for galaxies at stellar mass $M_* \sim 10^{9} M_\odot$ as a function of mass and redshift.  Values are calculated using the prescription given in \citet{Moster2011}, extrapolated to higher redshift as required.}
\label{fig:bias}
\end{figure}

\subsection{Clustering as a test of hierarchical merging}
The small volume probed by previous Frontier Field observations at $z\sim~8$ results in a very large expected field-to-field variance of $40-50$\%, currently dominated by Poisson error in each field at the bright end. This significantly limits the ability of the HFFs to measure accurate cosmic average quantities and true cosmic variance \citep{Robertson2014}.  The additional coverage from BUFFALO will reduce this field-to-field variance by $\sim2\times$, allowing an improved measurement of the true cosmic variance.

The true variance in number counts per field provides a direct and simple measure of the galaxy bias at these redshifts, allowing us to estimate the dark matter halo masses hosting these galaxies (e.g., \citealt{Adelberger1998,Robertson2010}). The availability of combined \emph{HST}+\emph{Spitzer} imaging will further allow us to determine the galaxy bias as a function of stellar mass - a crucial test of theoretical models \citep{Behroozi2015,OShea2015}.  Similar tests are not possible in the significantly more numerous BoRG (Brightest of Reionizing Galaxies, \citealt{Trenti2012}) fields because they lack ancillary \emph{Spitzer} data required to measure stellar masses of high-redshift galaxies, although followup \emph{Spitzer} observations have been conducted in some fields with high-redshift candidates \citep{Morishita2018,Bridge2019}.

\subsection{Improvement in Magnification Estimates}

As discussed in \S~\ref{sec:lensingmm}, a lack of knowledge of the presence of substructures in the outskirts of clusters can lead to underestimation of the mass and thus the density profile of the cluster by up to 20\% \citep{Acebron2017}. Magnification at a given location in the cluster is a direct product of the mass estimate (and the redshift of a background source). Therefore, the expanded BUFFALO observing area provides a unique dataset which can overcome the lack of accuracy of core-only mass models by being able to detect precisely where substructures are located and estimate precisely their mass \citep[down to a few percent precision thanks to \emph{HST} resolution,][]{A2744_HFF}. The exact improvement in magnification estimates with BUFFALO is difficult to predict as it strongly depends on the dynamical state of each clusters, i.e., how many substructures exist and how massive those substructures are. However, once all BUFFALO clusters are analyzed, we will have for the first time a 'statistical' sample allowing us to precisely estimate the bias induced by substructures on the magnification, which can then be related to the mass of those substructures.

\section{Structure evolution \& Dark Matter physics}
\label{sec:structure}

The BUFFALO survey offers an unprecedented look at the large-scale structure of our universe by providing high-resolution observations from space with \emph{HST} over an unusually wide area around the six massive Frontier Fields clusters. Flanking the core of those clusters with several pointings has extended the field of view out to $\sim$3/4 of the virial radius. This significant improvement is designed to allow improved investigations into the origin and evolution of large-scale structure.  The current consensus model, $\Lambda$CDM, assumes the hierarchical evolution of structures. They grow at a robustly-predicted rate, fed by the cosmic web's filaments which carry dark matter and baryons towards the centers of what will become large clusters \citep{Bond1996}.

However, so far, very few observations have been able to detect those large-scale filaments \citep{Dietrich2012,Jauzac2012,Eckert2015,Connor2018}. BUFFALO now allows an unprecedented resolution for mapping of the dark matter and will deliver data to detect material (both baryons and dark matter) falling on the clusters by the direct detection of galaxies and enhanced star-forming regions as tracers of the filaments and indirect detection of dark matter densities using weak lensing.  Clearly identifying and locating individual filaments and the local cosmic web could be useful in constraining their effect on star formation and quenching \citep{AragonCalvo2019}, galaxy angular momentum \citep{Goh2019} and even, through a "spiderweb test" \citep{Neyrinck2018}, a redshift-to-distance mapping.

In addition, reaching further away from the cluster center at the HST resolution will sharpen our understanding of merging events. Indeed, all the 6 HFF clusters show various states of merging (see \citet{Lotz2017} and references therein.)  The extended view of the clusters can now be used to map the underlying dark matter distribution using both strong and weak lensing. At the depth and width of existing Frontier Fields observations, strong lensing already revealed a detailed view of the dark matter distribution in the cluster cores, but it is known that the mass in cluster outskirts can influence the inner core profile \citep{Mahler2018,Acebron2017}.  BUFFALO now provides a unique view of the global mass distribution of clusters up to $\sim 3/4\,\times R_{\rm vir}$.

A few percent accuracy on weak-lensing measurements will be reachable due to highly constrained photometric redshift selection, primarly because of the unique \emph{HST} and \emph{Spitzer}/IRAC coverage of the BUFFALO fields.  Extensive spectroscopic follow-up of these clusters is planned as well.  The broader lensing mapping of the substructures will break the degeneracy coming from the outskirt mass contribution to the lensing potential. 

$\Lambda$CDM and hierarchical merging also predict the evolution of galaxies while falling in clusters. Extending \emph{HST} observations to $\sim$3/4\,$R_{\rm vir}$ also provide stringent observational constraints for the study of galaxy evolution within dense environments. 

\citet{Niemiec2018} studied and quantified dark matter stripping of galaxy halos during their infall into clusters, as well as the evolution of their stellar mass and star formation using the Illustris-1 simulation.
Such studies are often made in aggregate on a large number of clusters, but BUFFALO will do it on a cluster by cluster basis, avoiding smoothing out smaller scale influences on individual clusters. This also allows insight into the stripping of dark matter halos, environmental quenching mechanisms, including galaxy-galaxy interactions and harassment \citep{Moore1996}, strangulation \citep{Balogh2000,Peng2015}, the evolution of the stellar population, and the galaxy gas reservoir, providing critical insight into the cluster assembly history. The infall regions and intermediate-density environments traced by BUFFALO provide ideal observations to study these interactions \citep{Moss2006,Perez2009,Tonnesen2012}. 

\begin{figure*}
\begin{center}
\includegraphics[trim=4mm 12mm 9mm 20mm, clip,width=0.8\textwidth]{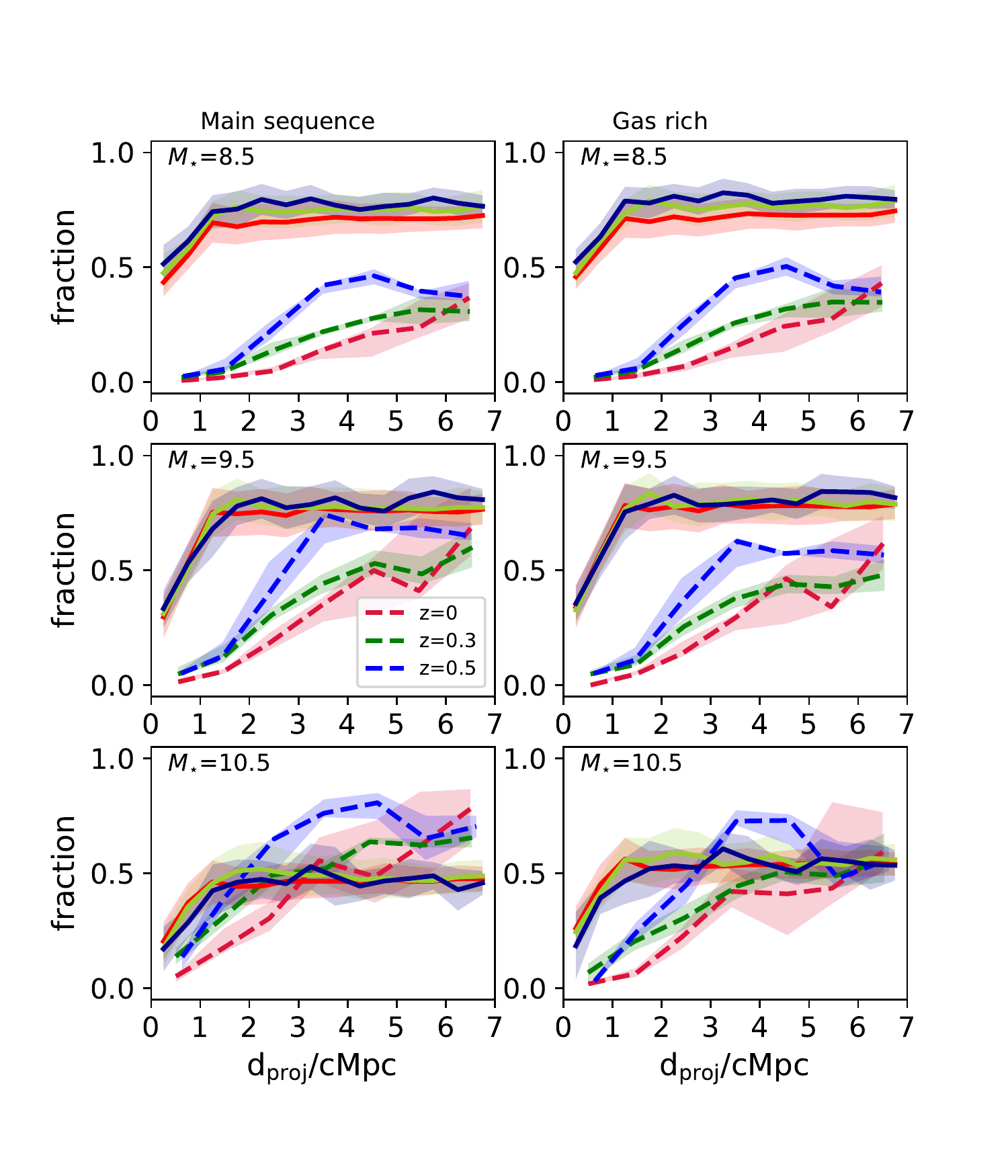}
\caption{The fraction of star-forming (left) and gas-rich (right) galaxies as a function of cluster-centric distance at three redshifts, $z=0,\ 0.3\ 0.5$ (red dotted, green dashed and blue solid lines, respectively), and three stellar mass bins, as labelled, in clusters of mass $M_{\rm 200c}>5\times 10^{14}\,\rm M_{\odot}$ in the cosmological semi-analytic model of galaxy formation {\sc Shark} \citep{Lagos2018} (solid lines) and the cluster hydrodynamical simulations zooms {\sc C-EAGLE} \citep{bahe2017,barnes2017} (dashed lines). For both simulations we show the medians and $25^{\rm th}-75^{\rm th}$ percentile ranges (thick lines with shaded regions, respectively). Star-forming and gas-rich galaxies are defined as those with $\rm sSFR/MS(M_{\rm star})>0.25$ and $f_{\rm neutral}/\langle f(M_{\rm star})\rangle > 0.25$, respectively. Here $\rm MS(M_{\rm star})$ and $f(M_{\rm star})$ are the main sequence sSFR and median $f_{\rm neutral}$ of main sequence galaxies at $M_{\rm star}$, respectively, and $f_{\rm neutral}=(M_{\rm HI}+M_{\rm H_2})/M_{\rm star}$.
}\mbox{}\label{figfractions}
\end{center}
\end{figure*}

\citet{Cen2014} report a steep increase in the fraction of star-forming galaxies from the cluster centre  up to $2\times R_{\rm vir}$. There is little consensus on this as it is shown in Fig.~\ref{figfractions} for two galaxy formation simulations, the semi-analytic model {\sc Shark} \citep{Lagos2018} and the cosmological hydrodynamical simulations cluster suite C-EAGLE \citep{bahe2017,barnes2017}. There are many differences worth highlighting - the simulations predict different star-forming/gas-rich galaxies overall fractions, particularly at low stellar masses ($<10^9\,\rm M_{\odot}$); C-EAGLE predicts fractions that continue to rise even out to $3-4\times R_{\rm vir}$, while {\sc Shark} predicts a steep increase out to $1-1.5\times R_{\rm vir}$ followed by a flattening; C-EAGLE predicts much stronger evolution over the last $5$ billion years than {\sc Shark}. This shows that the observations of BUFFALO combined with local Universe cluster observations will offer strong constraints on the models, hopefully allowing to rule out some of the wildly different behavior seen in Fig.~\ref{figfractions}. Note that these differences arise even though both simulations account for quenching mechanisms typical of galaxy clusters, such as ram pressure stripping, lack of cosmological accretion, among others \citep{Gunn1972,Bahe2013}. Fig.~\ref{figfractions} also shows the fraction of gas-rich galaxies, computed from their atomic plus molecular gas fraction, as a proxy for good candidates of ``jellyfish'' galaxies. \citet{Poggianti2018} show that the ``jellyfish'' galaxy population can be used to investigate the density of the cluster gas halos by probing the material in star-forming regions coming out of the galaxies as they fall into the cluster. In BUFFALO we expect to probe these galaxies at intermediate redshifts, building on previous studies in HFF clusters \citep{Schmidt2014,Treu2015,Vulcani2016}.

In recent decades, the theoretical framework that describes the formation of cosmic structures has been tested by increasingly precise observations \citep{Planck2016}, which show good agreement with several key aspects of current models. BUFFALO offers an orthogonal probe by providing a detailed view of the clustering as well as its synergy with simulations of the universe and massive clusters such as the HFF ones.  The Frontier Fields clusters remain rare massive clusters in simulations \citep{Jauzac2016,Jauzac2018} and a better understanding of their environment, only possible with BUFFALO, will shed light on how they form and evolve. By finding simulated analogs of those six clusters, it is also possible to test various related cosmological effects on the growth rate, mass ratio among substructures, and merging events.

\subsection{Cluster Science and Dark Matter Physics}
High-resolution space-based observations of galaxy clusters have revolutionized the study of dark matter \citep{Natarajan2002,A1689,Locuss_Richard}.  Early observations of merging galaxy clusters presented some of the most conclusive and unequivocal evidence for the existence of dark matter, while placing stringent limits on modified theories of gravity. 
For example the Bullet Cluster, a post-merger collision of two galaxy clusters in the plane of the sky, clearly shows the separation of intra-cluster gas and its associated cluster members \citep{bulletclusterA,bulletclusterB}. Thanks to weak-lensing, it was found that the majority of the mass lies in the galaxies, and not the gas, as suggested by modified gravity theories \citep{separation}. Following this study, multiple other merging clusters were found to exhibit similar properties, and detections of such offsets soon became ubiquitous \citep{minibullet,Merten2011,Harvey2015,Jauzac2016}. 

As a result of these studies it soon became clear that merging clusters could provide further insights into dark matter and not just evidence for its existence. For example, it is possible to constrain the self-interaction cross-section of dark matter using these clusters \citep{Harvey13,SIDMmodel,Robertson2019}. Self-interacting dark matter (SIDM) is an extension to the cold and collisionless dark matter paradigm that has received a lot of interest in the last decade. With its discriminative signals, tests of SIDM models provide a unique window in to the physics of the dark sector. 


With the discovery of multiple colliding systems it soon became possible to collate these events and statistically average them in order to provide further constraints on SIDM. Methods that estimated the relative positions of dark matter, galaxies and gas were developed to statistically average over many merging events \citep{bulleticity,HARVEY14}.  The first study of 30 merging galaxy clusters was carried out by \citet{Harvey2015}, and constrained the cross-section of dark matter to $\sigma_{\rm DM} /m< 0.5$cm$^2$/g. However, a subsequent study found potential systematic and revised this estimate to $\sigma_{\rm DM} /m< 2$cm$^2$/g \citep{Wittman2018}. Either way, it was clear that there was statistical potential in these merging systems.

The BUFFALO survey will extend the already successful HFF program, providing a unique insight into the dynamics of dark matter during halo in-fall. Probing the regions out to $\sim3/4 r_{\rm vir}$, BUFFALO will examine a regime where the unknowns of core-passage will be circumvented, and positional estimates of the substructures will be cleaner. 

Indeed, \emph{HST} high-resolution will allow us to precisely locate and weigh dark matter substructures in the clusters \citep[down to the percent level precision on mass measurement and down to 6\% on the location of dark matter peaks,][]{Jauzac2014MACSJ0416_HFF,A2744_HFF}.
It will then be possible to trace their dynamical history by combining the lensing analysis with the plethora of multi-wavelength data available on these six clusters, e.g., X-rays will help trace the gas, while optical, near-infrared, UV imaging and spectroscopy will help understand the dynamics and kinematics of the galaxies in the detected substructures.
Moreover, with sample-specific simulations of multiple dark matter models, it will be possible to monitor, test and mitigate all known and unknown systematic errors. As a result, BUFFALO will provide the cleanest measure of SIDM from in-falling substructures to-date. 

In addition to monitoring the trajectories of infalling halos to constrain SIDM, understanding the mass function within the clusters will provide important insights into the dynamics of dark matter. For example it has been suggested that A2744 exhibits too much substructure \citep{substructure_a2744A,Jauzac2016}, when compared to the predictions of standard CDM. This finding could possibly be an indication of exotic dark matter physics, however in such current small fields of view and on one sample, it is difficult to establish statistical significance. The extended imaging of BUFFALO with novel methods to compare observations to simulations such as the cluster power-spectrum \citep{substructureMF} and peak analysis \citep{peakAnal} will provide important insights in the substructure mass function of clusters and the dark matter that drives these statistics.

\subsection{3D halo structure and non-thermal pressure support}
\label{sect:3d}

How the dark and baryonic masses distribute in the halo's gravitational potential is a fundamental prediction of models of large scale structure formation,
allowing to use galaxy clusters as astrophysical laboratories, cosmological probes, and tests for fundamental physics. 
Probes at different wavelengths (optical, SZ, X-ray) can be used and combined to reconstruct three-dimensional triaxial ellipses describing the geometrical shape of the gas and total mass distribution, as obtained from the CLUster Multi-Probes in Three Dimensions (CLUMP-3D) project on 16 X-ray regular CLASH clusters
\citep{sereno18, umetsu18, chiu18}. In this analysis, weak lensing signal constrains the 2D mass and concentration which are deprojected thanks to the information on shape and orientation from X-ray (surface brightness and temperature) and SZ. The mass and concentration can be then determined together with the intrinsic shape and equilibrium status of the cluster as required by precision astronomy through a Bayesian inference method and not relying on the assumptions of spherical symmetry or hydrostatic equilibrium, which could bias results.
The joint exploitation of different data-sets improves the statistical accuracy and enables us to measure the 3D shape of the cluster's halo and any hydrostatic bias, evaluating the role of the non-thermal pressure support.
In general, they obtained that the shapes are in good agreement with the predictions from the standard $\Lambda$CDM cosmological model. 
However, compared to simulations, the data show a slight preference for more extreme minor-to-major axial ratios. We need a combination of more sensitive observational data to probe, also as function of halo mass and dynamical state, the 3D structure of the gas and dark matter distribution, assessing their consistency with $\Lambda$CDM predictions and their equilibrium once geometrical biases (like projection) are corrected for.

\subsection{Galaxy evolution}
 
As the largest observable gravitationally bound structures in the Universe,
galaxy clusters provide a unique tool for exploring the coeval evolution of
galaxies and cosmic structures. Very effective star formation quenching is
observed in clusters at all redshifts, where the fraction of star forming
galaxies is lower than in the field \citep{Hashimoto1998}, and the fraction of
early type morphologies (lenticulars, ellipticals) is the highest
\citep{Dressler1980}. A major thrust of ongoing research is to understand these transitions, particularly using deep HST imaging of clusters \citep{Martinet2017,Wagner2017,Marian2018,OlaveRojas2018,Connor2019,RodriguezMunoz2019}. 

There are strong hints that star formation suppression already occurs at large distances
from the cluster cores \citep{Haines2015, Bahe2015}, and that red galaxies are
located preferentially close to filament axes \citep[e.g.,][]{Malavasi2017,
Laigle2018}. Therefore, understanding pre-processing of galaxies requires
studies that move to regions well beyond the cluster core.
Color gradients are required to unveil the location of recent star-formation
\citep{Villalobos2012, Liu2018} as well as evidence for stripping and quenching,
e.g. so-called ``jellyfish" galaxies in the process of being ram-pressure
stripped \citep[e.g., in Abell\,2744, see][]{Owers2012}.

The BUFFALO survey is perfectly suited for these studies as it traces the
intermediate-density environments where the quenching processes occur. The
multi-wavelength data provides high-resolution measurements of the colors,
star-formation rates, morphologies and local environments of galaxies extending
out to the edges of the massive HFF clusters, enhancing the legacy value of the
Frontier Fields observations.

\subsection{Intracluster light}
\label{subsec:icl}

One of the most intriguing signatures of the assembly of galaxy clusters is a diffuse light known as intracluster light (ICL). This light is composed of a substantial fraction of stars not gravitationally bound to any particular galaxy but to the cluster potential. Observations have shown that the ICL is the product of interactions among the galaxies in the cluster. In this sense, the ICL is a unique tracer of how the assembly of the cluster proceeds through cosmic time \citep[see][for a review]{Montes2019b}.

The integrated stellar population properties revealed by the ICL tell us about the dominant process responsible of the formation of this diffuse light and therefore of the mechanisms at play in the assembly of the cluster. Different scenarios for the origin of the ICL result in different stellar population properties, ranging from the shredding of dwarf galaxies \citep{Purcell2007, Contini2014}, to violent mergers with the central galaxies of the cluster \citep{Murante2007, Conroy2007}, in situ formation \citep{Puchwein2010}, or the pre-processing of diffuse light of groups infalling into the cluster \citep{Mihos2004, DeLucia2012}. These different mechanisms might vary within the cluster and during the history of the cluster \citep{canas2019b}.  

Several works have already studied the ICL in the HFF clusters \citep{Montes2014, Montes2018, Jimenez-Teja2016, Morishita2017} but one of the difficulties encountered in these studies is the limited field of view of the HFF observations that prevent an accurate sky subtraction and therefore, accurate properties of this light at large radius from the central parts of the clusters. These studies have found that the main mechanism to produce ICL is the tidal stripping of massive satellites ($\sim 10^{10-11}$ M$_{\odot}$). However, these results only describe the more central parts of the clusters ($<200$ kpc).

\begin{figure*}
\begin{center}
\includegraphics[width=\textwidth,angle=0.0]{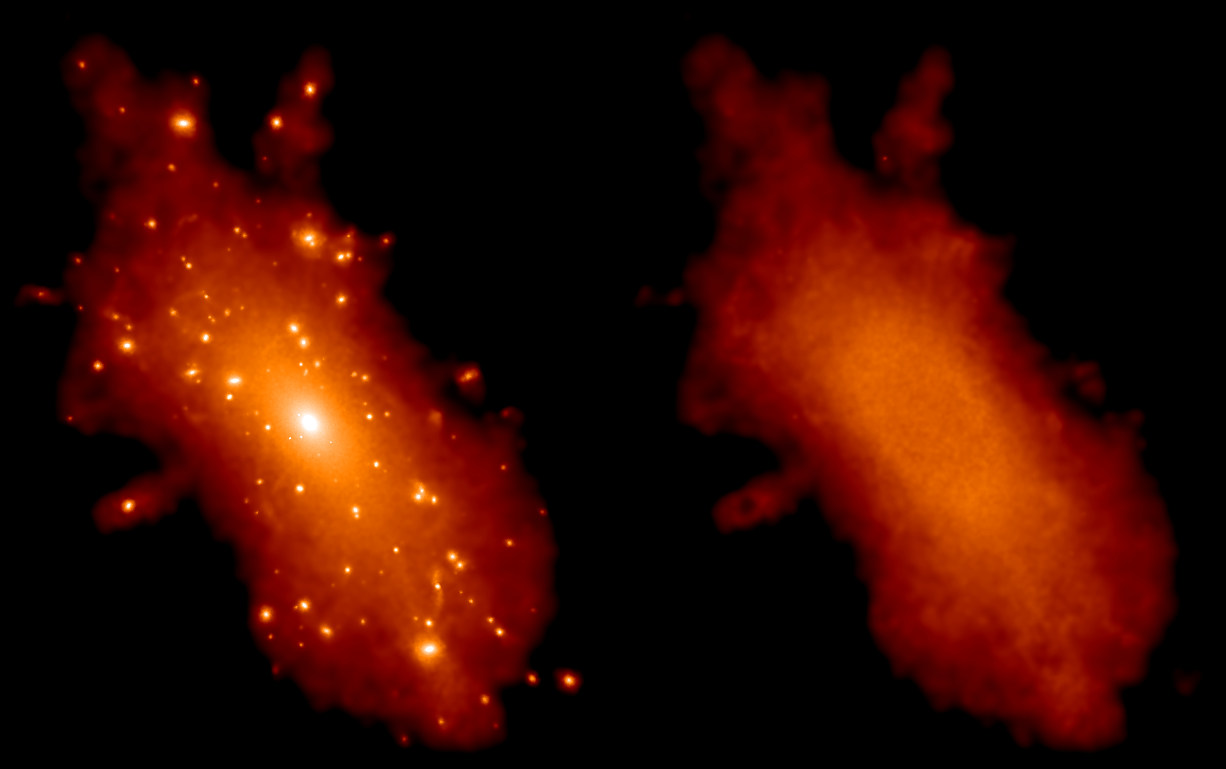}
\caption{Example of a simulated massive cluster of $M_{\rm 200,crit}\approx 10^{15}\,M_{\odot}$ from the C-EAGLE simulation \citep{bahe2017,barnes2017} at $z=0.3$. The left panel shows a stellar mass map of the whole 3D Friends-of-friends region while the right panel shows the ICL only, as defined as the kinematically hot stellar component using the full 6D information available in cosmological simulations (see \citealt{canas2019} for details of the algorithm and \citet{canas2019b} for an analysis of the intra-halo stellar component of galaxies across environments and cosmic time). BUFFALO will allow us to measure the fraction of total light in the ICL and its stellar population properties to test the predictions above and investigate how this component forms.
}\mbox{}\label{fig:ICL}
\end{center}
\end{figure*}

Using BUFFALO, we will be able to explore the ICL in detail up to $\sim 3/4$ $R_{vir}$. That will allow us to explore the formation mechanisms at play at large cluster radius and expand our knowledge of the formation of this diffuse light. 
The amount of light in the ICL provides information on the efficiency of the interactions that form this component (see Fig. \ref{fig:ICL}). To date, simulations have provided contradictory predictions as to how the amount of ICL depends on halo mass as well as the expected evolution with time, with some works finding no dependence on halo mass (e.g. \citealt{Contini2014, Rudick2011}) while some others report a clear dependence (e.g. \citealt{Murante2007, Cui2014}). Evolution wise, some authors report strong evolution of the ICL fraction (e.g. \citealt{Contini2014}) or rather weak (e.g. \citealt{Rudick2011}). Some of these discrepancies are due to the variety of numerical definitions of ICL (see discussion in \citealt{canas2019b}), which generally use the 3D position of particles in simulations. \citet{canas2019} introduced a full 6D method to define ICL and applied it to the Horizon-AGN simulations \citep{canas2019b} (see Fig.~\ref{fig:ICL}) and C-EAGLE (Ca\~nas et al. in preparation) to show that the ICL fraction is remarkably flat at the galaxy cluster regime, and that different simulations agree once the same ICL definition is applied. BUFFALO will place unique constraints on the ICL fraction, which together with local Universe measurements, will provide a large cosmic time baseline to compare with simulation predictions.
The multiwavelength coverage of BUFFALO is crucial to derive the properties of the stellar populations of the ICL within the cluster to study how its different formation mechanisms vary with clustercentric distance. We will also be able to measure diffuse light in substructures and quantify for the first time the amount of light that will end up as ICL through pre-processing. 

Finally, recent work has demonstrated that the ICL accurately follows the shape of the underlying dark matter halo in clusters of galaxies \citep{Montes2019}. This results was only possible thanks to the superb mass modelling available for the HFF clusters. The next step will be to extend this analysis to larger scales to assess whether the similarity between the distributions of ICL and mass holds at larger cluster radius. Simulations show that this similarity should hold out to 1.1 Mpc (Alonso-Asensio et al., in prep.), and with BUFFALO we will be able to finally assess it. 

One limiting factor for deriving ICL brightness profiles is establishing a sky background. The outskirts of the HFF primary field still contain contributions from ICL, but observations in the same filters in the parallel fields were taken at different times. Due to the temporal variations of the sky background, the parallel fields therefore cannot be used by themselves to calibrate a sky background for use in the cluster center \citep{DeMaio2015}.  

Thus, in addition to the previously-described observations, BUFFALO data of MACSJ1149 also include six additional orbits of imaging taken by GO-15308 (PI:A. Gonzalez). These six orbits were divided into three pointings of two orbits each; each pointing was observed with the F105W filter and the F160W filter on WFC3/IR for one orbit each. The three pointings linearly bridge the primary and parallel HFF fields such that each pointing has a 20\% spatial overlap with its neighbors. Due to this overlap, temporal variations can be accounted for, and the dominant uncertainty in sky background becomes that of the flat-fielding process. Based on work by \citet{DeMaio2018}, it is expected that the flat-fielding should reduce the residual systematic uncertainty to $\mu > 31\ \textrm{mag}\ \textrm{arcsec}^{-1}$, enabling a measurement of the ICL radial brightness profile down to $\mu_{160} = 29.5\ \textrm{mag}\ \textrm{arcsec}^{-1}$. In addition, GO-15308 includes parallel ACS observations in  F606W and F814W, which extend the coverage of MACSJ1149 in three pointings per filter to the side of the cluster opposite the  HFF parallel field.

\section{Supernovae and Other Transients}
\label{sec:supernovae}
The observations of this program have been purposely spaced into two visits (or epochs) per pointing, to allow for the discovery of supernovae (SNe) and other transients. By design, revisits to the same field in the same filter sets are separated by approximately 30 days, approximately matching the rise time of most SN types, including type Ia SNe, around $z=1$. We expect this to yield a discovery rate per visit that is comparable to that in the Cluster Lensing And Supernovae with Hubble (CLASH; \citealt{Postman2012}), which will amount to about 10 to 25 events over the full duration of the BUFFALO survey. 

\begin{figure}[ht]
\begin{center}
    \includegraphics[width=0.47\textwidth]{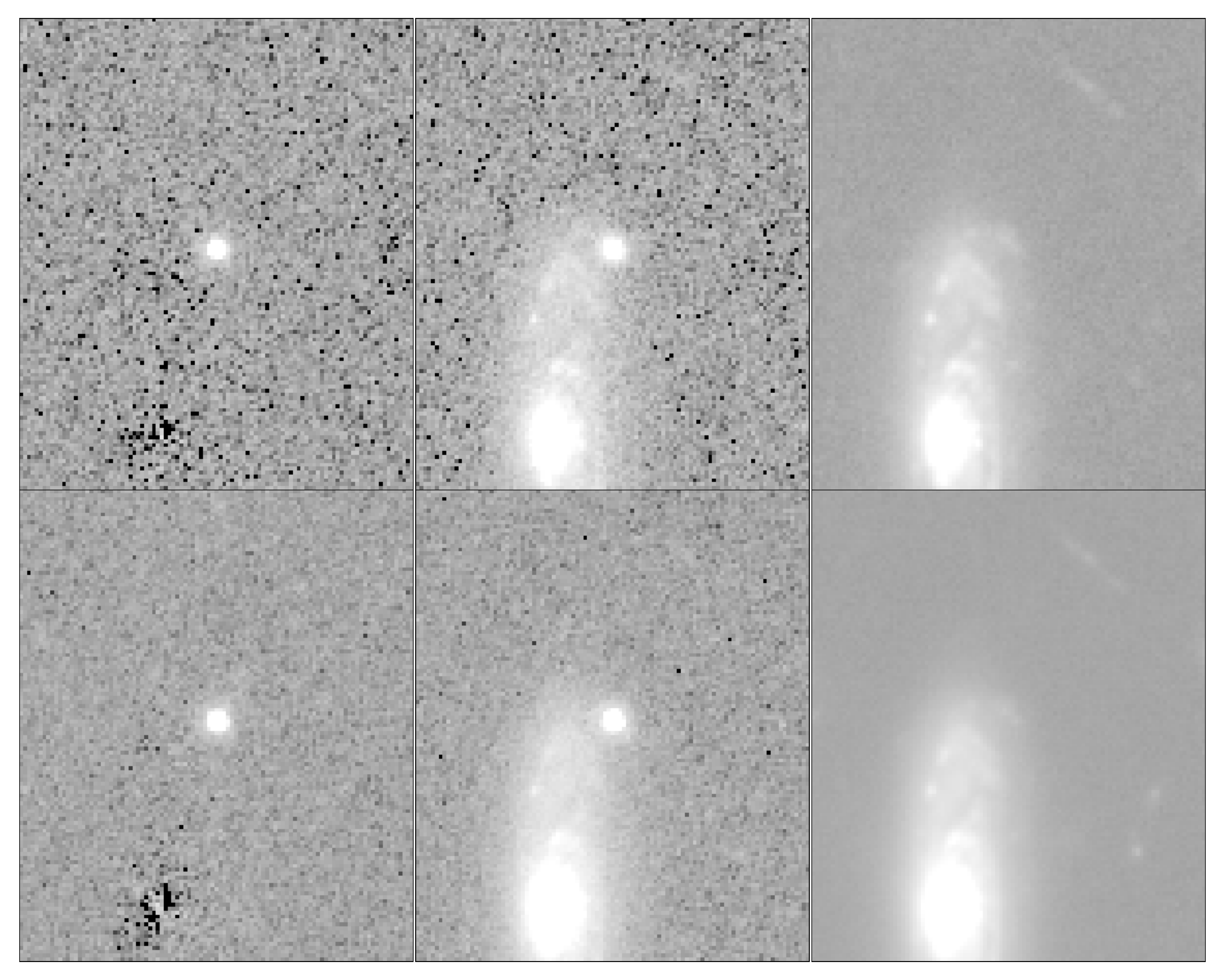}
    \caption{Example of supernova detected in the BUFFALO fields at right ascension 22:49:13.700, declination -44:32:38.736. From right to left are 6 arcsecond cutouts of the reference HFF image, the BUFFALO epoch 1 image, and the difference image. Top row shows the F606W filter and the bottom row shows the F814W filter.}
\end{center}
\label{fig:sn_detect_example}
\end{figure}

An example of one such discovery is given in Figure \ref{fig:sn_detect_example}, a transient discovered in the AbellS1063 difference imaging at 22:49:13.70, -44:32:38.74. The cutouts show a 6 arcsecond box around the region, showing the reference HFF image, first epoch BUFFALO observation, and the difference frame in which the search for supernovae was conducted. This event is visible in both the ACS observations of the parallel field, taken April and May 2019, but also the first epoch of the WFC3 parallel pointing, taken October 2019, highlighting the power of the various time delays between BUFFALO images in the search for transient events.

A summary of the preliminary searches of each field, conducted after each epoch was processed, for supernovae and transients is shown in Table \ref{tab:sne-candidates}. As of the 22nd of October 21 of the 28 camera-pointing combinations have been conducted, and thus the preliminary findings of 6 SNe suggests a total of 8 over the entire BUFFALO program, slightly lower than predicted by CLASH. However, more thorough follow up searches will be conducted which may yield fainter or more obscure detections, increasing the detection rate.

\begin{figure}[ht]
\begin{center}
\includegraphics[width=0.47\textwidth]{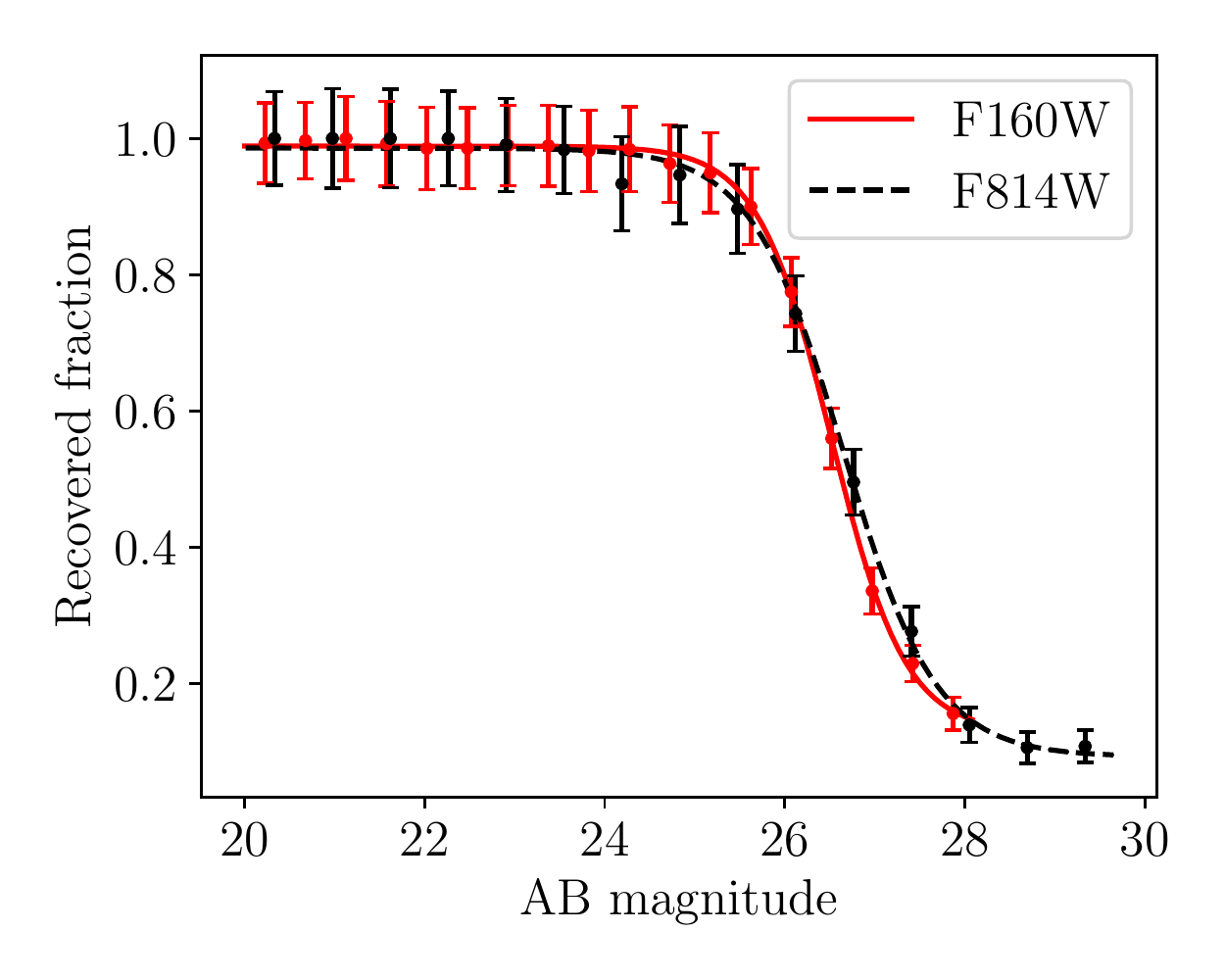}
\caption{Expected efficiency of supernovae searches within the BUFFALO images. The recovered fraction of fake sources injected into BUFFALO difference images as a function of magnitude for the two reddest filters of the two \textit{HST} cameras -- F814W for ACS, black dashed line, and F160W for WFC3/IR, red solid line -- is shown. Sources are recovered if the peak pixel count rate of the source is above 0.03 count/s. The 50\% recovery rate is 26.5 mag for both filters.}

\end{center}
\label{fig:sn_recover_rate}
\end{figure}

To probe the expected sensitivity of the BUFFALO difference images, we conducted a simple test. Fake sources were injected randomly into the drizzled difference images, with AB magnitudes in the range 20 to $\sim$30, with appropriate Poisson noise. These were then recovered using a simple peak finding algorithm, looking for sources with peak count rates above 0.03 count/s -- roughly the lowest pixel count rate that is qualitatively visible above the BUFFALO difference images background by eye. The recovery fraction as a function of magnitude is shown in Figure \ref{fig:sn_recover_rate}. Given the similar zero points of the two filters, the 50\% recovery rate is 26.5 mag and the $\sim$95\% recovery rate is $\sim$25th magnitude for both. The very faintest recoverable SNe are 27-28th magnitude in the F160W filter, and 29-30th magnitude in F814W, due in part to the differing detector pixel scales, drizzled onto a common 0.06 arcsecond scale. However, it should be noted that this efficiency rate is quite optimistic, as it requires a fairly quiescient background in both frames contributing to the difference image. High background count rates, or supernovae lying near the center of galaxies, will suffer higher backgrounds or poor dither pattern noise (see left panels, Figure \ref{fig:sn_detect_example}), raising the minimum count rate at which transients are detectable. Therefore a more conservative estimate might put the minimum count rate at 0.3 count/s, and thus reduce the 50\% completeness limit of detections to 24th magnitude, and the 95\% recovery rate to 22.5 mag. As all currently detected sources are of order 21-22nd magnitude (see Table \ref{tab:sne-candidates}), it is likely that we are complete at this brighter detection threshold, but will require follow up studies to recover the very faintest objects.

 \begin{deluxetable*}{ccccccccc}
 \tablecaption{\label{tab:sne-candidates}
	BUFFALO Supernovae Candidates}
\tablehead{%
Field	        	& RA (H:M:S) & Dec (D:M:S)	& Detected Filters	& AB  Mag & Date & Pointing    & Epoch   & Redshift       }
\startdata
MACS0416 & 04:16:14.25 & -24:03:41.16 & F606W & $\sim$22-23 & Feb 2019 & Main & Epoch 2 & z $\sim$ 0.3 \\
 & 04:16:33.05 & -24:06:44.66 & F606W & & Sep 2019 & Parallel & Epoch 2 & \\
 AbellS1063 & 22:48:53.56 & -44:31:19.59 & F105W & & Jun 2019 & Main & Epoch 2 & \\
  & & & F125W & & & & & \\
  & & & F160W & & & & & \\
  & 22:49:13.70 & -44:32:38.74 & F606W & 22.40$\pm$0.20 & Jun 2019 & Parallel & Epoch 2 & \\
  & & & F814W & 21.80$\pm$0.20 & & & & \\
  & & & F105W & 22.69$\pm$0.03 & Oct 2019 & Parallel & Epoch 1 & \\
  & & & F125W & 22.69$\pm$0.03 & & & & \\
  & & & F160W & 22.61$\pm$0.04 & & & & \\
 Abell 2744 & 00:14:28.55 & -30:23:33.98 & F606W & & Jul 2019 & Main & Epoch 2 & z $\sim$ 0.2-0.3 \\
  & & & F814W & & & & & \\
  & 0:14:26.56 & -30:23:44.17 & F606W & & Jul 2019 & Main & Epoch 2 & \\
  & & & F814W & & & & & \\
\enddata
\end{deluxetable*}

The BUFFALO observations may contribute SN detections at a uniquely high redshift range.   Infrared imaging programs with HST like this one have been the most efficient approach for discovery of SNe at $1.5<z<2.5$, the highest redshift regime where SNe Ia and normal luminosity core collapse SNe (CCSNe) have been detected  \citep{Graur2014,Rodney2014,Rodney2015b,Strolger2015}.  Gravitational lensing from the clusters that dominate the center of each BUFFALO field will magnify background SNe, making it possible to detect SNe at redshifts $2 < z < 3$ that would normally be undetectable (cf. \citealt{Rubin2018}), though lensing does reduce the high-z survey volume behind the cluster \citep{Barbary2012}.

The BUFFALO program also could potentially locate SNe behind the clusters that are significantly magnified by gravitational lensing \citep{Patel2014,Nordin2014,Rodney2015a}. These lens-magnified events can provide a valuable test of the gravitational lensing models, and have  been informative in the refinement of lens models for Frontier Fields clusters. 
There is also a small, but non-zero, chance of locating another strongly-lensed SN with multiple images, 
similar to ``SN Refsdal'' \citep{Kelly2015}.  A SN that is multiply-imaged by a cluster lens is likely to have measurable time delays \citep{Kelly2016,Rodney2016}, and could potentially be used to measure the Hubble constant \citep{Grillo2018}. Deep imaging surveys of 
the Frontier Fields clusters have revealed other exotic transient events, with extreme 
magnifications $\mu>100$ \citep{Rodney2018} or even $\mu>1000$ \citep{Kelly2018,Chen2019}
The BUFFALO imaging could discover such events at redshifts $0.8\lesssim z\lesssim2$, but the rates 
of such extreme microlensing events are highly uncertain, and the BUFFALO cadence is not 
optimized for their discovery (see, however,
\citet{Kaurov2019} in which they predict the interval between caustic transient
events of ~1 yr).

\section{Data Products}
\label{sec:dataproducts}

A major goal of BUFFALO is to produce data products which can be used by the entire astronomical community, both as standalone catalogs and in preparation for additional observations with the \emph{James Webb Space Telescope (JWST)}.  Release of these products will be via the Mikulski Archive for Space Telescopes\footnote{http://archive.stsci.edu/}.  In an effort to release initial data as quickly as possible, these catalogs will be released individually for each of the six Frontier Fields clusters and parallel fields rather than waiting for the entire program to be complete.   

\subsection{High-level Mosaics}

\begin{figure*}
\includegraphics[width=\textwidth]{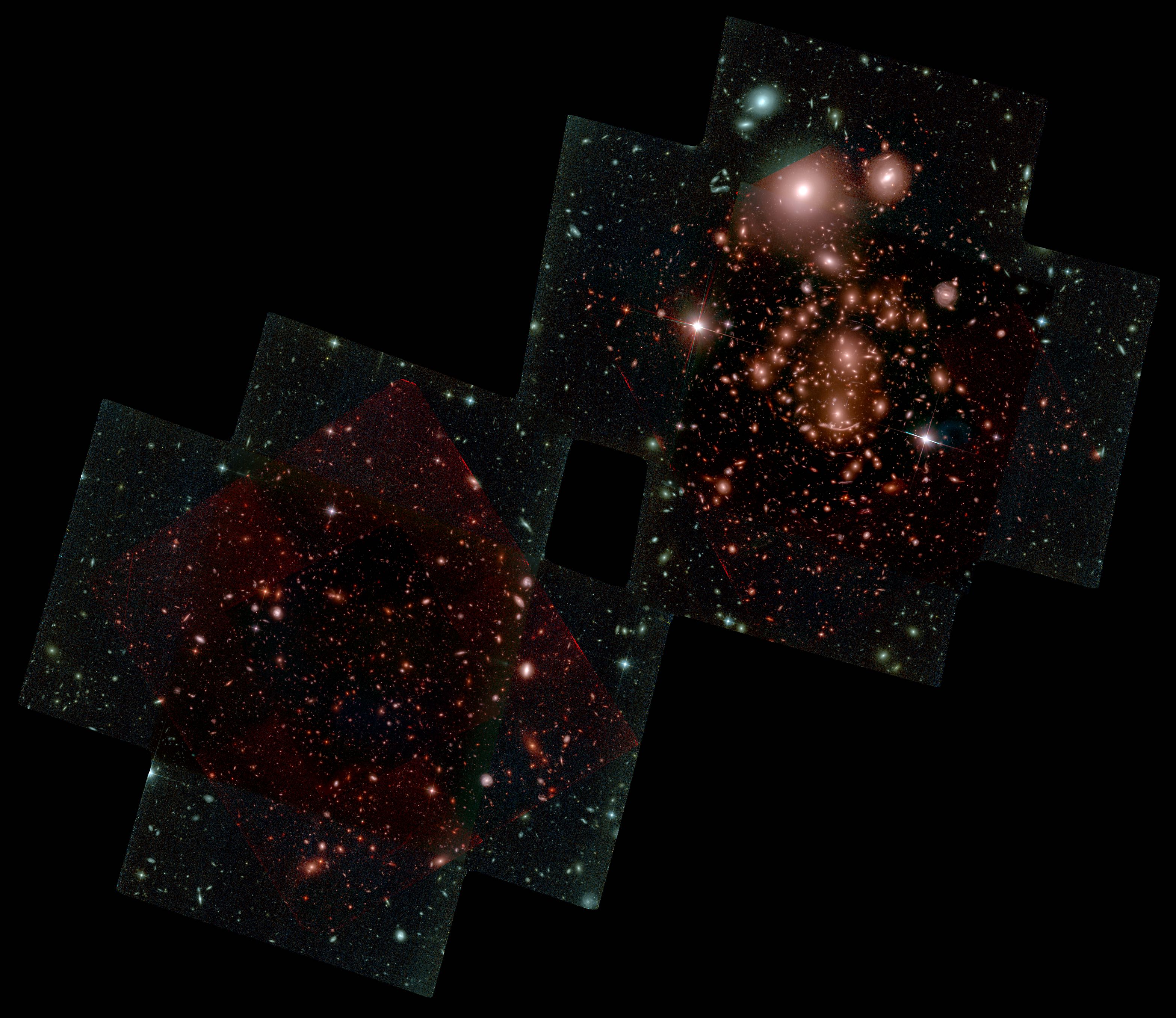}
\caption{BUFFALO composite color image of Abell 370.  The BUFFALO field of view is four times larger than the previous Frontier Fields coverage (shaded in the central region for both the cluster and parallel fields) in addition to increasing the depth in the central region.}
\label{fig:mosaic}
\end{figure*}

BUFFALO will release a mosaic in each of the five HST bands, combining the new data with previous Frontier Fields observations in the same bands where it is available.  The first of these mosaics, Abell 370 (Fig. \ref{fig:mosaic}, is now available on MAST concurrent with the publication of this paper.  Mosaics for the other clusters will be released as available, with a strong effort to release as many as possible prior to relevant proposal deadlines.  Although these first data will be quickly presented, it is anticipated that more significant scientific value will come from value-added, multi-wavelength catalogs and associated models.  

\subsection{Catalogs}

BUFFALO will produce several value-added catalogs, with an initial release for Abell 370 and updated planned as additional clusters are completed.  These catalogs will exploit BUFFALO photometric data along with all available multi-wavelength observations from the HST archive. A list of existing datasets can be found in Table \ref{tab:otherdata}.

It will also include ancillary data from near-IR surveys (\citealt{KIFF}; Nonino et al., in prep.\footnote{http://archive.eso.org/cms/eso-archive-news/first-data-release-from-the-galaxy-clusters-at-vircam-gcav-eso-vista-public-survey.html}) and the \emph{Spitzer} images that are a cornerstone of the BUFFALO program (see \S~\ref{sec:observations}). Photometric redshifts will be derived from such a large photometric baseline by means of state-of-the-art software as EAZY \citep{EAZY} and LePhare \citep{Arnouts1999,Ilbert2006}, fitting galaxy (and stellar) templates to the spectral energy distribution (SED) of each object. 
Galaxy physical properties (stellar mass, star formation rate, etc.) will be inferred by means of an additional SED fitting phase based on stellar population synthesis models \citep{Conroy2013}. 
Recent studies have found the inclusion of nebular emission to be crucial to estimating physical parameters of galaxies from SED-fitting, in particular at high redshift \citep[see e.g.][]{schaerer12}. An additional method combining stellar population models \citep{bruzual03} with nebular emission line models \citep{gutkin16} in an SED-fit with the Bayesian code \texttt{BEAGLE} \citep{chevallard16}, which is optimized to yield both photometric redshifts and galaxy physical parameters simultaneously, will therefore also be applied to the catalogs.
Alternate ``data-driven'' methods not relying on synthetic templates (cf.\ \citealt{vanderMaaten2008,vanderMaaten2014}; Steinhardt et al., in prep.) will be also applied to show the potential of novel machine learning methods in this field of research. 
The BUFFALO team includes authors of reference SED fitting studies and techniques as well as HFF luminosity and mass functions \citep{EAZY,Coe2015,Ishigaki2015,Connor2017,Davidzon2017,Ishigaki2017,Kawamata2018} and will use this knowledge to produce a comprehensive ``consensus catalog''  designed to be used for a wide range of analyses. 
Moreover, structural parameters for foreground galaxies 
will be provided via morphological analysis through GALFIT \citep{galfit3} and GALAPAGOS \citep{galapagos}. These codes will be applied with a strategy similar to that used in \citet{Morishita2017}. 

The Abell 370 catalogs will be released in Pagul et al.\ and Niemiec et al. (both in preparation), including:.   

\begin{itemize}
    \item Photometry
    \item Photometric redshift (plus spectroscopic redshifts when available) and physical properties
    \item Cluster membership
    \item ICL map and structural parameters of foreground galaxies 
    \item Mass models as described in \S~\ref{subsec:massmodeling}
\end{itemize}

We refer the reader to those papers for further details about the catalog-making process.  Catalogs for the other clusters are expected to be released approximately six months after their observations are completed (Table \ref{tab:obs-schedule}) and will include the same products as the Abell 370 catalog.

\subsection{Mass Models}
\label{subsec:massmodeling}

The BUFFALO team includes groups which have been responsible for producing a variety of independent mass models for the HFF clusters from earlier datasets \citep{Jauzac2014MACSJ0416_HFF,Johnson2014,Lam2014,Richard2014,Diego2015b,Diego2015,A2744_HFF,Jauzac2015b,Oguri2015,MACSJ1149_HFF_cats,Wang2015,Williams2015,Diego2016b,Diego2016,Harvey2016,Jauzac2016Refsdal,Johnson2016,Kawamata2016,Sebesta2016,Treu2016,Priewe2017,Diego2018,Finney2018,Jauzac2018,Mahler2018,Strait2018,Williams2018,Sebesta2019,Williams2019}.  Each of these groups will generate an independent mass model (e.g., Niemiec et al. in prep. for Abell 370), based upon a variety of fitting methods (cf. \S~\ref{sec:lensingmm}). In order to analyse the intrinsic properties of background lensed galaxies, the BUFFALO collaboration will follow the \textit{HFF} philosophy and release high-level data products through the MAST archive for the scientific community. An effort is being made to release all of the models for each specific cluster simultaneously in order to encourage a comparison between them. 

For strong, weak or joint strong and weak lensing modelings, each independent team from the BUFFALO collaboration will deliver mass, magnification at several redshift, deflection, convergence and shear maps together with their associated error maps, as well as weak-lensing catalogues. Finally, an online magnification calculator will also be available for fast magnification and errors estimates.

\subsection{Planned Releases}

Data products for each cluster will be released individually, in order to provide the initial data as rapidly as possible.  The first release will contain a mosaic, followed later by the value-added catalogs and mass models.  If HST observations follow the current schedule (\S~\ref{sec:observations}), this would result in approximately the release schedule shown in Table \ref{tab:rel-schedule}.

 \begin{deluxetable*}{lccc}[t]
\tablecaption{\label{tab:rel-schedule}
	BUFFALO Planned Data Releases}
\tablehead{%
Field	        	& Observations Complete	& Mosaic	& Catalogs and Mass Models   }
\startdata
Abell 370           & 2019 Jan		& 2019 Dec & 2020 Jan	\\
					&		    &   	&   	        	\\
MACS J0717.5+3745	& 2019 Apr		& 2019 Dec & 2020 Apr	\\
					&	    	&   	&   				        	\\
MACS J0416.1-2403	& 2019 Sep		& 2020 Mar & 2020 Sep	\\
					&	    	&   	&   				        	\\
Abell S1063		    & 2019 Nov		& 2020 May & 2020 Nov	\\
					&	    	&	    &   					        \\
Abell 2744		    & 2019 Dec		& 2020 Jun & 2020 Dec	\\
					&	    	&   	&   			        	\\
MACS J1149.5+2223	& 2020 May		& 2020 Nov & 2021 May	\\
\enddata
\end{deluxetable*}

\section{Summary}
\label{sec:summary}

The BUFFALO survey will expand the area of \emph{HST} coverage by approximately a factor of four around the six Hubble Frontier Fields (HFF), a region which already has multi-wavelength coverage including ultradeep \emph{Spitzer} imaging.  BUFFALO covers this region in five filters, WFC3/IR F105W, F125W, and F160W along with ACS/WFC F606W and F814W, with depths chosen based upon what has been learned from existing surveys both in HFF and other ultradeep fields.  As with the original HFF program, this expanded coverage will simultaneously provide new insights into a wide range of problems at both high and low redshift.  

The expanded coverage will not only discover many new sources at $z>7$, with the highest-redshift BUFFALO source most likely to lie at $z \sim 9-10$, but also provide a significant improvement in measurements of cosmic variance.  Both will be important for designing observational programs with \emph{JWST}.  The former is necessary because the NIRSPEC field of view is larger than previous HFF coverage, but fits within BUFFALO.  The latter will be important both for designing \emph{JWST} surveys and as a test of theoretical models of early galaxy assembly.

The same observations will allow an improvement in the mass models of these clusters, both in their central regions and in cluster outskirts.  BUFFALO is to be the first large \emph{HST} program with an emphasis on studying the dynamics of infalling cluster substructures.  Filamentary structures may contain even a majority of the mass and provide critical insights into the dynamics of galaxy assembly and the cosmic web, and these studies will likely be a significant part of the legacy value of BUFFALO.  At the same time, the improvement in mass models of the clusters themselves will not only improve our understanding of structure evolution and dark matter physics, but also improve magnification maps and therefore our existing measurements of the highest-redshift galaxy population accessible prior to \emph{JWST}.

Finally, BUFFALO holds the possibility for serendipitous discoveries.  The most probable would be the discovery of a lensed supernova or other lensed high-redshift transient.  If one is found, it would allow a major improvement in mass models and related studies for the cluster in which it is found.  Because of increased cosmic variance towards high redshift and towards high mass, it is also possible that BUFFALO will discover a very early galaxy or assembling galaxy protocluster. 

The BUFFALO survey is currently scheduled to have observations completed in May 2020.  Data will be released individually as cluster analysis is completed, with the first data from Abell 370 available on MAST concurrent with the publication of this paper.  Mosaics will be available on MAST first, with value-added catalogs and a variety of mass and lensing models made available later. 

\acknowledgements

CS acknowledges support from the ERC Consolidator Grant funding scheme (project ConTExT, grant No. 648179). The Cosmic Dawn Center is funded by the Danish National Research Foundation under grant No. 140.  MJ is supported by the United Kingdom Research and Innovation (UKRI) Future Leaders Fellowship 'Using Cosmic Beasts to uncover the Nature of Dark Matter' [grant number MR/S017216/1]. This project was also supported by the Science and Technology Facilities Council [grant number ST/L00075X/1].
Joseph Allingham is supported by the Postgraduate Research Scholarship in Astroparticle Physics/Cosmology in the University of Sydney.  AA acknowledges the support of Royal Society through Wolfson Fellowship program.   YMB acknowledges funding from the EU Horizon 2020 research and innovation programme under Marie Sk{\l}odowska-Curie grant agreement 747645 (ClusterGal) and the Netherlands Organisation for Scientific Research (NWO) through VENI grant 016.183.011.  RJB gratefully acknowledges the support of TOP grant TOP1.16.057.  RC is supported by the MERAC foundation postdoctoral grant awarded to CL, and by the Consejo Nacional de Ciencia y Tecnolog\'ia CONACYT CVU 520137 Scholar 290609 Overseas Scholarship 438594.  The work of TC was carried out at the Jet Propulsion Laboratory, California Institute of Technology, under a contract with NASA.  B.D. acknowledges financial support from NASA through the Astrophysics Data Analysis Program (ADAP), grant number NNX12AE20G, and the National Science Foundation, grant number 1716907.  JMD. acknowledges the support of project PGC2018-101814-B-100 (MCIU/AEI/MINECO/FEDER, UE) Ministerio de Ciencia, investigaci\'on y Universidades.  A.C.E. acknowledges support from STFC grant ST/P00541/1.  S.E. acknowledges financial contribution from the contracts ASI 2015-046-R.0 and ASI-INAF n.2017-14-H.0, and from INAF ''Call per interventi aggiuntivi a sostegno della ricerca di main stream di INAF''.  O.G. is supported by an NSF Astronomy and Astrophysics Fellowship under award AST-1602595.  DH acknowledges the support by the Delta ITP foundation.  EJ acknowledges AMU, CNRS and CNES for their support   CL is funded by the ARC Centre of Excellence for All Sky Astrophysics in 3 Dimensions (ASTRO 3D), through project number CE170100013.   ML acknowledges CNRS and CNES for support  GEM acknowledges support from the Villum Fonden  research  grant  13160  “Gas  to  stars,  stars  todust:  tracing  star  formation  across  cosmic  time”.  The work of L.A.M. took place at the Jet Propulsion Laboratory, California Institute of Technology, under a contract with NASA.  J.R. was supported by JPL, which is run under a contract for NASA by Caltech.  RMR acknowledges support from HST-GO-15117.005  MS is supported by the Netherlands Organization for Scientific Research (NWO) VENI grant 639.041.749  HYS acknowledges the support from NSFC of China under grant 11973070 and the Shanghai Committee of Science and Technology grant No.19ZR1466600.  GPS acknowledges financial support from STFC grant number ST/N000633/1.  CT acknowledges the support of the Deutsche Forschungsgemeinschaft under BA 1369/28-1  ST acknowledges support from the ERC Consolidator Grant funding scheme (project ConTExT, grant No. 648179). The Cosmic Dawn Center is funded by the Danish National Research Foundation under grant No. 140.  K.U. was supported by the Ministry of Science and Technology of Taiwan (grant MOST 106-2628-M-001-003-MY3) and by the Academia Sinica Investigator Award (grant AS-IA-107-M01).   RJvW acknowledges support from the VIDI research programme with project number 639.042.729, which is financed by the Netherlands Organisation for Scientific Research (NWO).  This work has made use of the CANDIDE Cluster at the Institut d'Astrophysique de Paris and made possible by grants from the PNCG and the DIM-ACAV.
\bibliographystyle{aasjournal}
\bibliography{ref}

\pagebreak

\begin{longrotatetable}
\begin{deluxetable*}{lccccc}
\startlongtable
\tablecaption{\label{tab:otherdata}
	Existing Multi-wavelength Frontier Fields Coverage}

\tablehead{%
Field	        	& Observatory	& Wavelengths	& Depth & Reference }
\startdata
Abell 370           & VLT/HAWK-I   & 2.2$\mu m$  & $\sim$ 26.18  &  \citet{KIFF}	\\
          & Spitzer IRAC 1,2  & 3.6$\mu m$, 4.5$\mu m$ & $\sim$ 25.19, 25.09  & (PI: T. Soifer
and P. Capak)	\\
          & Spitzer IRAC 3,4 (cluster-only)  & 5.8$\mu m$, 8.0$\mu m$ & $\sim$ 23.94, 23.39    &	\\
          & Spitzer MIPS (cluster-only)  & 24$\mu m$ & $\sim$ 17.88    &	\\
         	& Chandra (X-ray)	&  515 &  88.0 	\\
					& XMM-Newton (X-ray)	&  0782150101 &  133.0 	\\
					          		& Bolocam			&  140 GHz 													& 11.8h 					& \citet{say+al13} \\
					& Planck				&  100, 143, 217, 353, 545, 857 GHz 								& -- 					& \citet{planck_2015_XXVII} \\
										& Subaru/Suprime-Cam	&  $B_\textrm{J}$,$R_\textrm{C}$,$I_\textrm{C}$,$i^\prime$,$z^\prime$ 						&  3.24ks ($R_\textrm{C}$))		& \citet{wtg_I_14,ser15_comalit_III} \\
					&		    &   	&   	        \\
					MACS J0717.5+3745	& Keck/MOSFIRE   & 2.2$\mu m$  & $\sim$ 25.31   &	\citet{KIFF}\\
	& Spitzer IRAC 1,2,3,4   & 3.5$\mu m$, 4.5 $\mu m$  &  $\sim$ 25.04, 25.17  &(PI: T. Soifer
and P. Capak)	\\
& Spitzer IRAC 3,4   & 5.8$\mu m$, 8.0$\mu m$  &  $\sim$ 23.94, 23.39  &	\\
	& Spitzer MIPS   & 24$\mu m$  &  $\sim$ 17.35   &	\\
	& Chandra (X-ray)	&  4200 &  58.5	 &  \citet{donahue14} \\
					& Chandra (X-ray)	&  1655, 16235, 16305 &  19.9, 70.1, 94.3	 & \citet{jauzac18} \\
					& XMM-Newton  (X-ray)	& 0672420101,   0672420201, 0672420301 &  61.2, 69.3, 64.1	 &  \citet{jauzac18} \\
					&		    &   	&   	        	\\
MACS J0416.1-2403	& VLT/HAWK-I   & 2.2$\mu m$  & $\sim$ 26.25   & \citet{KIFF}	\\
	& Spitzer   & 3.5$\mu m$, 4.5 $\mu m$  & $\sim$ 25.31, 25.44    &	(PI: T. Soifer
and P. Capak)\\        	      	
		& Chandra (X-ray)	&  10446 &  15.8 	&  \citet{donahue14} \\
					& Chandra (X-ray)	&  16236, 16237, 16523 &  39.9, 36.6, 71.1	&  \citet{balestra16, bonamigo18} \\
					& Chandra (X-ray)	&  16304, 17313 &  97.8, 62.8	&  \citet{balestra16, bonamigo18} \\
                	& Bolocam			&  140 GHz 													& 7.8h 						& \citet{say+al13}\\
					& Planck				&  100, 143, 217, 353, 545, 857 GHz 								& -- 					& \citet{planck_2015_XXVII}\\
					& Subaru/Suprime-Cam	&  $B_\textrm{J}$,$R_\textrm{C}$,$z^\prime$ 									&  --						& \citet{ume+al14,ser15_comalit_III}\\
					&	    	&   &	&     \\
Abell S1063		    & VLT/HAWK-I   & 2.2$\mu m$  & $\sim$ 26.31   & \citet{KIFF} \\
   & Spitzer IRAC 1,2  & 3.6$\mu m$, 4.5$\mu m$ & $\sim$ 25.04, 25.04     &(PI: T. Soifer
and P. Capak)	\\
   & Spitzer IRAC 3,4 (cluster-only)  & 5.8$\mu m$, 8.0$\mu m$   & $\sim$ 22.96, 22.64  &   &	\\ & Spitzer MIPS (cluster-only)  & 24$\mu m$   & $\sim$ 18.33   &	\\		       
	    & Chandra (X-ray)	&  4966, 3595 &  26.7, 19.9 	&  \citet{donahue14} \\
					& Chandra (X-ray)	&  18611, 18818 &  49.5, 47.5 	&  \citet{bonamigo18} \\
					& XMM-Newton  (X-ray)	&  0504630101   &   	52.7  &    \citet{lovisari17}  \\
	    	& Planck				&  100, 143, 217, 353, 545, 857 GHz 								& -- 						& \citet{planck_2015_XXVII}\\
					& ESO/WFI			&  $U_{877}$,$B_{842}$,$V_{843}$,$R_{844}$,$I_{879}$,$z_{846}$ 		&  --						& \citet{ume+al14,ser15_comalit_III}\\
					&		    &   	&   	        	\\
Abell 2744		    & VLT/HAWK-I   & 2.2$\mu m$  & $\sim$ 26.28  &  \citet{KIFF} \\
	    & Spitzer IRAC 1,2  & 3.6$\mu m$, 4.5$\mu m$ & $\sim$ 25.32, 25.08  &   (PI: T. Soifer
and P. Capak)	\\
        & Spitzer IRAC 3,4 (cluster-only)  & 5.8$\mu m$, 8.0$\mu m$   & $\sim$ 22.78, 22.45    &	\\
        & Spitzer MIPS (cluster-only)  & 24$\mu m$   & $\sim$ 18.23    &	\\
					& XMM-Newton  (X-ray)	&  0743850101  & 111.9 & \citet{eckert15} \\
		    	& Bolocam			&  140 GHz 													& -- 						& \citet{say+al16}	\\
					& Planck				&  100, 143, 217, 353, 545, 857 GHz 								& -- 						& \citet{planck_2015_XXVII}\\
					& Subaru/Suprime-Cam	&  $B_\textrm{J}$,$R_\textrm{C}$,$i^\prime$,$z^\prime$ 								&  3.12ks ($R_\textrm{C}$)		& \citet{med+al16,ser15_comalit_III}	\\
					&	    	&   	&   			        	\\
MACS J1149.5+2223	& Keck/MOSFIRE   & 2.2$\mu m$  & $\sim$ 25.41   & \citet{KIFF} \\
	& Spitzer   &  3.5$\mu m$, 4.5 $\mu m$ &  $\sim$ 25.24, 25.01 &  (PI: T. Soifer
and P. Capak)	\\
	& Chandra (X-ray)	&  3589 &  20.0 	&  \citet{donahue14} \\
					& Chandra (X-ray)	&  1656, 16238, 16239, 16306 &  18.5, 35.6, 51.4, 79.7	& \\
					& Chandra (X-ray)	&  16582, 17595, 17596  &  18.8,  69.2, 72.1	& \\
					& XMM-Newton  (X-ray)	&  0693661701  & 28.9 & \citet{lovisari17} \\
            	& Bolocam			&  140 GHz 													& 17.7h 					& \citet{say+al13}\\
					& Planck				&  100, 143, 217, 353, 545, 857 GHz 								& -- 						& \citet{planck_2015_XXVII}\\
					& Subaru/Suprime-Cam	&  $B_\textrm{J}$,$V_\textrm{J}$,$R_\textrm{C}$,$I_\textrm{C}$,$i^\prime$,$z^\prime$ 			&  1.94ks ($I_\textrm{C}$))		& \citet{ume+al14}; \\
					& & & & \citet{wtg_I_14,ser15_comalit_III}
\enddata
\end{deluxetable*}
\end{longrotatetable}

\end{document}